%% file: main.tex
\documentclass[acmsmall,nonacm,screen]{acmart}

\usepackage{graphicx}
\usepackage{multirow}
\usepackage{makecell}
\usepackage{cleveref}
\usepackage{tabularx}
\usepackage{ragged2e}
\usepackage{csquotes}
\usepackage{pifont}
\newcommand{\cmark}{\ding{51}}%
\newcommand{\xmark}{\ding{55}}%
\usepackage{listings}
\usepackage{paralist}
\usepackage{tabularray}
\usepackage{subcaption}
\usepackage{enumitem}
\usepackage[linewidth=1pt]{mdframed}

\newcommand\mrev{}

\AtBeginDocument{%
  \providecommand\BibTeX{{%
    \normalfont B\kern-0.5em{\scshape i\kern-0.25em b}\kern-0.8em\TeX}}}

\setcopyright{acmlicensed}
\copyrightyear{2024}
\acmVolume{}
\acmNumber{}
\acmArticle{}
\acmYear{2024}
\acmMonth{4}

\acmJournal{CSUR}

\begin{document}

\title{Warm-Starting and Quantum Computing: A Systematic Mapping Study}

\author{Felix Truger}
\email{truger@iaas.uni-stuttgart.de}
\orcid{0000-0001-6587-6431}
\author{Johanna Barzen}
\email{barzen@iaas.uni-stuttgart.de}
\orcid{0000-0001-8397-7973}
\author{Marvin Bechtold}
\email{bechtold@iaas.uni-stuttgart.de}
\orcid{0000-0002-7770-7296}
\author{Martin Beisel}
\email{beisel@iaas.uni-stuttgart.de}
\orcid{0000-0003-2617-751X}
\author{Frank Leymann}
\email{leymann@iaas.uni-stuttgart.de}
\orcid{0000-0002-9123-259X}
\author{Alexander Mandl}
\email{mandl@iaas.uni-stuttgart.de}
\orcid{0000-0003-4502-6119}
\author{Vladimir Yussupov}
\email{yussupov@iaas.uni-stuttgart.de}
\orcid{0000-0002-6498-637X}

\affiliation{%
  \institution{Institute of Architecture of Application Systems, University of Stuttgart}
  \streetaddress{Universit\"atsstraße 38}
  \city{70569 Stuttgart}
  \country{Germany}
}

\renewcommand{\shortauthors}{Truger et al.}

\begin{abstract}
  \input{content/abstract}
\end{abstract}

\ccsdesc[500]{General and reference~Surveys and overviews}
\ccsdesc[500]{Computer systems organization~Quantum computing}
\ccsdesc[500]{Software and its engineering}
\ccsdesc[100]{Hardware~Quantum computation}

\keywords{Warm-Start, Quantum Software Engineering, Quantum Algorithm, Systematic Mapping Study.}

\authorsaddresses{\vspace{-2.5mm}\begin{mdframed}\large{© Truger et al., 2024. This is the authors' version of the work. It is posted here for your personal use. Not for redistribution. The definitive version was published in ACM Computing Surveys, \url{https://doi.org/10.1145/3652510}.}\end{mdframed}
\vspace{1mm}
Authors’ address: 
\href{https://orcid.org/0000-0001-6587-6431}{Felix Truger}, truger@iaas.uni-stuttgart.de; 
\href{https://orcid.org/0000-0001-8397-7973}{Johanna Barzen}, barzen@iaas.uni-stuttgart.de; 
\href{https://orcid.org/0000-0002-7770-7296}{Marvin Bechtold}, bechtold@iaas.uni-stuttgart.de; 
\href{https://orcid.org/0000-0003-2617-751X}{Martin Beisel}, beisel@iaas.uni-stuttgart.de; 
\href{https://orcid.org/0000-0002-9123-259X}{Frank Leymann}, leymann@iaas.uni-stuttgart.de;
\href{https://orcid.org/0000-0003-4502-6119}{Alexander Mandl}, mandl@iaas.uni-stuttgart.de; 
\href{https://orcid.org/0000-0002-6498-637X}{Vladimir Yussupov}, yussupov@iaas.uni-stuttgart.de, 
Institute of Architecture of Application Systems, University of Stuttgart, Universitätsstraße 38, 70569 Stuttgart, Germany.}
\maketitle

\input{content/Introduction}
\input{content/BackgroundMotivation}

\input{content/ResearchDesign}
\input{content/Results}
\input{content/Limitations}
\input{content/RelatedWork}
\input{content/Conclusion}

\section*{Data Availability and Reproducibility}
The data of this study including the search queries and intermediate results is published in a separate data repository~\cite{datarepo} to enable transparency, traceability, and full reproducibility of the results.\looseness=-1
\begin{acks}
This work was partially funded by the project \textit{SEQUOIA} funded by the Ministry of Economic Affairs, Labour and Tourism of Baden-W\"urttemberg and by the BMWK projects \textit{SeQuenC}~(01MQ22009B), \textit{EniQmA}~(01MQ22007B), and \textit{PlanQK}~(01MK20005N).
\end{acks}

\bibliographystyle{definitions/ACM-Reference-Format}
\bibliography{bibliography}

\appendix
\input{content/Appendix}

\end{document}

%% file: content/abstract.tex


Due to low numbers of qubits and their error-proneness, Noisy Intermediate-Scale Quantum (NISQ) computers impose constraints on the size of quantum algorithms they can successfully execute.
State-of-the-art research introduces various techniques addressing these limitations by utilizing known or inexpensively generated approximations, solutions, or models as a starting point to approach a task instead of starting from scratch.
These so-called warm-starting techniques aim to reduce quantum resource consumption, thus facilitating the design of algorithms suiting the capabilities of NISQ computers.
In this work, we collect and analyze scientific literature on warm-starting techniques in the quantum computing domain.
In particular, we
(i) create a systematic map of state-of-the-art research on warm-starting techniques using established guidelines for systematic mapping studies, 
(ii) identify relevant properties of such techniques, and 
(iii) based on these properties classify the techniques identified in the literature in an extensible classification scheme.
Our results provide insights into the research field and aim to help quantum software engineers to categorize warm-starting techniques and apply them in practice.
Moreover, our contributions may serve as a starting point for further research on the warm-starting topic since they provide an overview of existing work and facilitate the identification of research gaps.\looseness=-1

%% file: content/Introduction.tex

\section{Introduction}
\label{sec:intro}
Quantum computers promise to solve a number of problems that are intractable on classical hardware~\cite{Preskill2018quantumcomputingin,gheorghiu2019verification}.
However, contemporary Noisy Intermediate-Scale Quantum~(NISQ) devices are error-prone and impose restrictions on the depth of quantum circuits they can successfully execute~\cite{leymann2020bitter,Preskill2018quantumcomputingin}.
Such limitations and the scarcity of contemporary NISQ devices encourage the development of specialized quantum algorithms and techniques, e.g., the Quantum Approximate Optimization Algorithm~(QAOA) is able to address the circuit depth limitation by means of an adaptable circuit depth~\cite{farhi2014quantum}.
Furthermore, quantum computation on NISQ devices is currently possible only via cloud services from a limited number of vendors~\cite{Leymann2020_QuantumCloud}, thus, computational tasks on NISQ devices can be either scheduled via reserved time slots or queued for execution on the respective cloud offering~\cite{LaRose2019overviewcomparison}.
In both cases, quantum devices may not be available as needed or only after an undesirable, significant waiting time.
Hence, to increase the efficiency of quantum applications and reduce waiting times, the utilization of quantum devices should be optimized.
One class of techniques addressing the mentioned issues is frequently termed as \textit{warm-starting} with various approaches falling into this class, e.g., pre-processing using widely available classical hardware or utilization of previous (quantum) computation results to improve the accuracy and speed of the computation.
\looseness=-1

The term \enquote{warm-starting} is widely used in classical computation, e.g., in machine learning and optimization, to describe techniques that reduce the usage of resources~\cite{kumar2020efficient,ash2020warm,sokoler2013warm,wang2016warm,benson2008interior,poloczek2016warm}.
The rough idea is to utilize known or inexpensively generated approximations, solutions, or models as a starting point to improve upon instead of starting from scratch.
Another category of warm-starting techniques in classical computation focuses on the preparation of an execution environment, e.g., reusing a running container in Function-as-a-Service offerings, as opposed to cold-starting a new container~\cite{Manner2018coldstart,yussupov2019smsFaaS}.
Although multiple research works focus on warm-starting in the domain of classical computation~\cite{yildirim2002warm,ash2020warm}, the concept appears to be still lacking a broader understanding in the quantum computing domain --
while a number of publications~\cite{egger2021warm,wurtz2021fixed,mari2020transfer,Weigold2021_HybridPatterns} focus on the topic of warm-starting quantum algorithms, there are no comprehensive secondary studies aiming to categorize such approaches.
In particular, many of these warm-starting techniques in the quantum computing domain differ significantly from each other in various properties.
For example, they target a quantum algorithm in different ways, such as biasing the initial quantum state as opposed to selecting advantageous initial circuit parameters in VQAs.
Furthermore, different warm-starting techniques may be applicable in conjunction with each other, however, approaches on how to categorize them and check their compatibility with each other are currently missing.
Therefore, a categorization of warm-starting techniques and an overview of their properties would be beneficial to researchers and quantum software engineers to understand which techniques, or combinations thereof, suit their quantum applications.

In this work, we aim to provide a systematic overview of existing, state-of-the-art warm-starting techniques employed in the domain of quantum computing by means of a systematic mapping study~(SMS).
The study relies on established guidelines for SMSs~\cite{petersen2008guidelines,petersen2015guidelinesUpdate}:
The multi-phase literature search and selection process comprises database searches, snowballing, and precise selection criteria to identify relevant publications.
Six well-known electronic databases for research publications in software engineering were queried for relevant publications, including the ACM Digital Library, IEEE Xplore, ScienceDirect, and SpringerLink.
After snowballing and selection, a final set of $80$ relevant publications served as the basis for the mapping study as well as the classification of warm-starting techniques in the quantum computing domain.
The contributions of this work are as follows:\looseness=-1
\begin{compactenum}[(i)]
    \item We derive and document a reusable search strategy that helps to identify publications related to warm-starting in the quantum computing domain,
    \item we create a systematic map of state-of-the-art research on the topic,
    \item we identify relevant properties of different kinds of warm-starting techniques, and
    \item based on these properties and the identified literature, we propose a classification of warm-starting techniques in the context of quantum algorithms.
\end{compactenum}

The remainder of this paper is structured as follows: 
In~\Cref{sec:background}, we provide the necessary background and motivate our work.
\Cref{sec:researchmethod} presents the research design in detail, including the search strategy and analysis concept.
The results are presented and discussed in~\Cref{sec:results}.
In~\Cref{sec:limitations}, we elaborate on the limitations and threats to validity of the study.
\Cref{sec:relatedwork}~presents related work before the paper is concluded with~\Cref{sec:conclusion}.

%% file: content/BackgroundMotivation.tex

\section{Quantum Algorithms and Warm-Starting: Background and Motivation}
\label{sec:background}
\label{sec:motivation}
The landscape of quantum computing evolves quickly: 
NISQ devices with few tens to hundreds of quantum bits~(qubits) are already available via cloud services~\cite{Leymann2020_QuantumCloud}, e.g., provided by IBM\footnote{\url{https://quantum-computing.ibm.com/}}, Amazon\footnote{\url{https://aws.amazon.com/braket/}}, Google\footnote{\url{https://cloud.google.com/}}, and others.
In addition to their relatively low numbers of qubits, these devices are error-prone, i.e., the control over their qubits is imprecise and qubit states suffer from instability~\cite{leymann2020bitter,Preskill2018quantumcomputingin}.
Gate-based, universal NISQ devices are currently deemed the most promising type of quantum computers compared to other quantum hardware, such as analog quantum simulators and specialized quantum annealers~\cite{Preskill2018quantumcomputingin}.
Quantum algorithms for gate-based NISQ devices are implemented as quantum circuits, which comprise layers of operations represented by different quantum gates, that are simultaneously applied to the qubits to manipulate their states and, thereby, perform a computation.
Quantum gates are a major source of errors.
Due to imperfect hardware, the underlying operations are applied with small deviations.
In deep circuits, i.e., circuits with many layers of gates, such gate errors will accumulate throughout the computation. 
Additionally, since qubit states are volatile, longer execution time makes them more likely to decay.
Consequently, deep circuits may lead to unusable results upon measurement.
The instability of qubits and erroneous gate operations, therefore, effectively limit the depth of executable quantum circuits on NISQ devices.
\looseness=-1

\mrev{
Thus, various promising algorithms, such as Shor's algorithm for integer factorization~\cite{shor1994factoring}, cannot be executed on current NISQ devices.
Since the development of fault-tolerant devices capable of executing these algorithms may still take many years~\cite{Preskill2018quantumcomputingin}, if feasible at all~\cite{dyakonov2007fault}, much of the current research focuses on quantum algorithms for NISQ devices that manage to contain the circuit depth.
Such algorithms are often Variational Quantum Algorithms~(VQAs) based on shallow Parameterized Quantum Circuits~(PQCs), i.e., quantum circuits comprising gates tunable by parameter values~\cite{cerezo2021variational}.
VQAs are hybrid quantum-classical algorithms that involve a loop in which an optimization algorithm executed on classical hardware is concerned with improving parameter values for a PQC such that the result quality increases.
During this optimization, the PQC is executed repeatedly with the continuously updated set of parameter values to derive directions for their improvement, e.g., by gradient descent.
Prominent examples of VQAs include the above-mentioned QAOA, Variational Quantum Eigensolvers~(VQEs), and Quantum Neural Networks~(QNNs).
\looseness=-1}

\mrev{To further boost the capabilities of NISQ-compatible quantum algorithms, various improvements have been proposed, e.g., to reduce the resource requirements or increase the quality of solutions~\cite{egger2021warm,tate2020bridging,galda2021transferability,mari2020transfer,Bechtold2023_CircuitCuttingQAOA}.
One group of techniques is frequently labeled as warm-starts.
For instance, warm-starts can be realized by encoding information obtained from classical approximation algorithms into the initial state of a quantum circuit.
Warm-started variants of the QAOA for the Maximum Cut problem (MaxCut) initialized with approximations have shown potential in improving the solution quality~\cite{tate2020bridging,egger2021warm}.
Other forms of warm-starts focus on viable parameter initializations for VQAs to reduce the resource requirements of the optimization.
For example, it was shown that optimized QAOA parameters correlate for related instances of MaxCut and can thus be transferred as initial parameters from one instance to another to substantially accelerate the parameter search~\cite{galda2021transferability}.\looseness=-1}

\mrev{In general, t}he term \textit{warm-starting} is widely used in various domains and represents heterogeneous techniques aiming to improve the efficiency of executing a certain target entity.
While combustion engines are warm-started in a rather literal sense~\cite{liu2013cold}, the term is used figuratively for computer systems~\cite{lin2019mitigating} and algorithms~\cite{ash2020warm, sokoler2013warm, wang2016warm, benson2008interior}.
In the context of classical algorithms, warm-starting is often described as the initialization of an optimization or training task with known, previous, or related solutions rather than starting from scratch~\cite{kumar2020efficient,ash2020warm,baylor2017tfx,sokoler2013warm,wang2016warm,benson2008interior,karumbaiah2021using}.
Warm-starting can be found, for example, in the fields of neural network training~\cite{kumar2020efficient,ash2020warm}, optimization~\cite{sokoler2013warm,wang2016warm,benson2008interior,poloczek2016warm,karumbaiah2021using}, and automated algorithm configuration~\cite{lindauer2018warmstarting}.
Often, the term \textit{cold-starting} is used to refer to cases where no warm-starting is employed~\cite{sokoler2013warm,wang2016warm,benson2008interior}, e.g., for the performance validation of warm-starting techniques.
The main goals aimed at with warm-starting techniques for algorithms include the reduction of  time~\cite{hartlep2015nmpc,baylor2017tfx,sokoler2013warm,wang2016warm,benson2008interior,poloczek2016warm,ralphs2006duality} and resource consumption~\cite{ash2020warm,han2020battery}.
Warm-starting particularly excels in optimization, since the search for an optimum often depends heavily on a good starting point.
Thus, such a starting point can help to find the optimum more efficiently.
However, it has also been found, that warm-starting can harm performance in some cases, e.g., the generalization of deep neural networks~\cite{ash2020warm}.
\looseness=-1

\begin{figure}[t]
    \centering
    \includegraphics[page=4,width=0.7\linewidth,trim=0 0 135 240, clip]{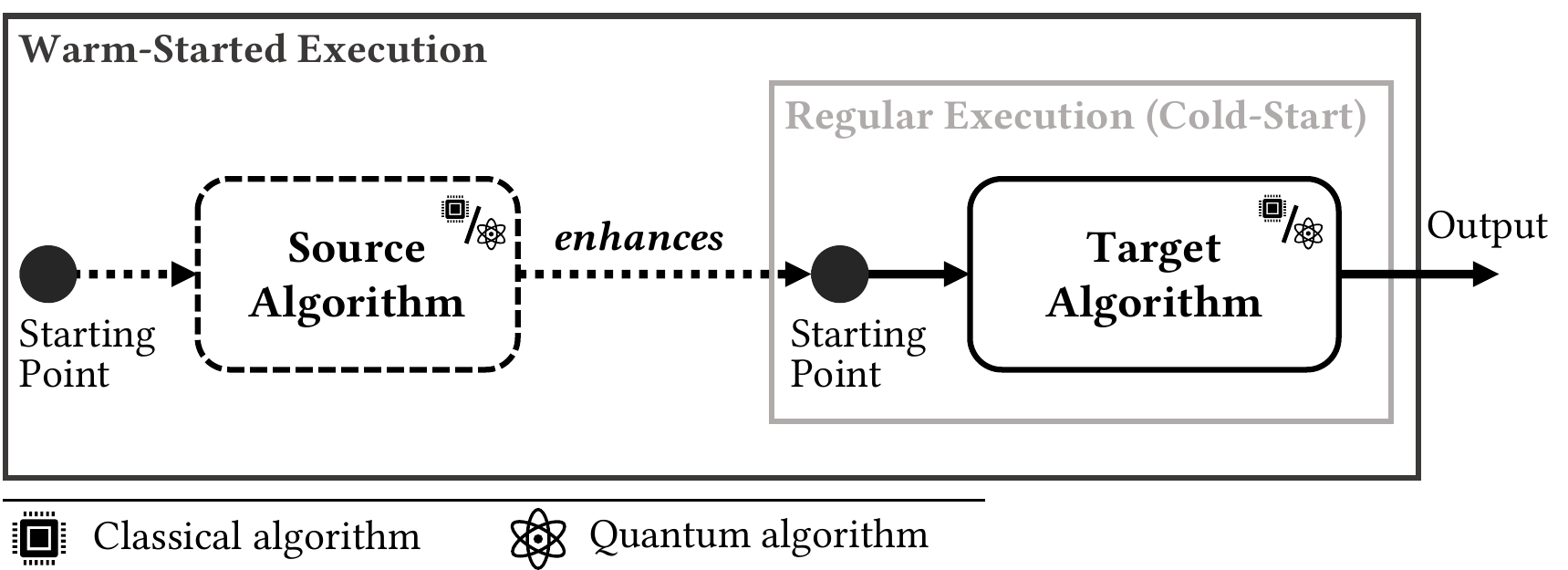}
    \caption{A high-level view on warm-started execution (black) and cold-started execution (gray) of algorithms.}
    \label{fig:warm_start_definition}
\end{figure}
From a high-level perspective, warm-starting in the context of quantum algorithms can be described as a two-stage process similar to how warm-staring in the classical computing domain has been described~\cite{gondzio1998warm}. 
As shown in \Cref{fig:warm_start_definition}, it consists of two algorithms that are executed sequentially: 
(i) A \textit{target algorithm} designed to solve a problem of interest and
(ii) a \textit{source algorithm} that \textit{enhances} the target algorithm in some way, e.g., by manipulating the inputs of the target algorithm or an environment that it runs in.
For warm-starting techniques in the context of quantum computing, at least one of the two algorithms is a quantum algorithm.
\mrev{
Since most quantum algorithms are hybrid~\cite{leymann2020bitter}, the term \enquote{quantum algorithm} in this work also refers to hybrid quantum-classical algorithms.}
The way algorithms are combined is analogous to classical warm-starting: The source algorithm is executed prior to the target algorithm, enhancing it in a certain intended manner, e.g., to achieve faster convergence, lower computational power consumption, or increased accuracy compared to cold-started executions.

Due to the anticipated heterogeneity of entry points for warm-starting techniques in the quantum computing domain, \mrev{e.g., in hybrid QNN training~\cite{qi2021classical,mari2020transfer}, variational parameter optimization~\cite{wurtz2021fixed,galda2021transferability}, and initial states of quantum algorithms~\cite{egger2021warm,tate2020bridging},} quantum software engineers need to understand the respective mechanisms and requirements to enhance their quantum applications.
The first challenge for understanding the warm-starting techniques and requirements lies in the identification of types of warm-starting techniques used in the context of quantum algorithms, which in turn requires a strategy for identifying relevant publications.
However, not all relevant publications explicitly mention “warm-starting” as a term, but rather describe approaches involving characteristic elements of warm-starts, such as,
\begin{compactitem}
    \item \textit{\enquote{pre-computation}, \enquote{pre-training}, or \enquote{transfer learning}} for neural networks,
    \item \textit{\enquote{initial parameters}, \enquote{parameter transfers}, or \enquote{pre-optimization}} for variational parameter optimization, or 
    \item \textit{\enquote{initialization strategies}} for encoding pre-computed values into the initial state of quantum circuits.
\end{compactitem}
Such a large variety of terms that may indicate warm-starting techniques implies the need for a sophisticated and systematic search strategy to identify as many relevant publications as possible.

The variety and complexity of warm-starting techniques makes them hard to identify not only because of different terminology, but also because they need to be distinguished from mere sequences of processing steps, that do not make use of warm-starts.
For example, Shor's algorithm for factorization~\cite{shor1994factoring} consists of classical parts run before and after a quantum circuit is executed on a quantum device~\cite{barzen2022continued}.
However, this composition of classical and quantum algorithms cannot be considered as warm-starting, since all processing steps are vital for the algorithm.
In contrast, the target algorithm of a warm-starting technique needs to be capable of an independent cold-start, i.e., execution in a standalone fashion to solve the problem of interest~(see~\Cref{fig:warm_start_definition}).

To the best of our knowledge, no attempt to analyze and categorize available warm-starting techniques exists, neither for classical-only nor for quantum-specific contexts.
Therefore, to address this research gap in the context of warm-starting techniques involving quantum algorithms, in this work, we aim to answer the following question: \textit{\enquote{What are the currently available warm-starting techniques, their properties, and related publication trends in the context of quantum algorithms?}}

%% file: content/ResearchDesign.tex

\section{Research Design}
\label{sec:researchmethod}
To pursue the question posed in~\Cref{sec:motivation}, this study has three consecutive goals, namely, (i) to create a systematic map of existing research focusing on warm-starting techniques that involve quantum algorithms, (ii) to analyze these techniques, and (iii) to categorize them.
To achieve this, we follow the guidelines for SMSs in software engineering~\cite{petersen2008guidelines,petersen2015guidelinesUpdate,kitchenham2007guidelinesSLR}.
This section introduces the research design in detail.
First, we formulate precise research questions before presenting the individual steps for the literature search.
Afterward, we describe the data extraction and analysis process.
The dataset of this study, including details of the literature search, the search results, extracted information, and intermediate results, is made available in a separate data repository~\cite{datarepo}.
\looseness=-1

\subsection{Research Questions}
\label{subsec:rqs}
We formulate the following three research questions for this study:
\looseness=-1

\smallskip\noindent
\textbf{Research Question 1 (RQ1):} \textit{\enquote{What are the publication trends of research literature focusing on and involving warm-starting techniques in the context of quantum algorithms?}}
By answering this question, we capture the current state-of-the-art of research on warm-starting techniques in the context of quantum algorithms.
The publication trends allow us to determine the growth of the research area over time.
By tracking different publication venues, we can also observe how the popularity of different types of venues changed over time. 
Further, we aim to identify the major research communities by tracking the affiliations of authors publishing in the field.

\smallskip\noindent
\textbf{Research Question 2 (RQ2):} \textit{\enquote{Which computational problems and quantum algorithms are addressed using warm-starting techniques and what are the benefits of warm-starting?}}
With this question, we deepen the analysis of the research trends to identify problems and concrete quantum algorithms that researchers have so far focused on with warm-starting techniques.
This also helps identify research gaps, since an overview of the existing research can be easily reviewed by researchers to find whether and which warm-starting techniques have already been applied in the context of a certain quantum algorithm.
Further, we capture the potential benefits of warm-starting techniques in the context of quantum algorithms as stated in the identified research literature.
An overview over the benefits may motivate employing and further developing such techniques.

\smallskip\noindent
\textbf{Research Question 3 (RQ3):} \textit{\enquote{Which properties characterize warm-starting techniques and how can different warm-starting techniques be classified based on these properties?}}
By answering this question, we identify relevant properties of warm-starting techniques based on which they can be classified.
Such a classification will provide a comprehensive overview of different warm-starting techniques and facilitate the application of warm-starting in different scenarios, e.g., for quantum algorithms that have not benefited from warm-starts yet.
Further, it will simplify combining warm-starting techniques, since the overview of properties makes it easier to comprehend the compatibility of different techniques.

\subsection{Search Strategy}
\label{subsec:search_strategy}
\begin{figure}[h]
    \centering
    \includegraphics[page=2,width=0.90\linewidth,trim=0 0 101 380,clip]{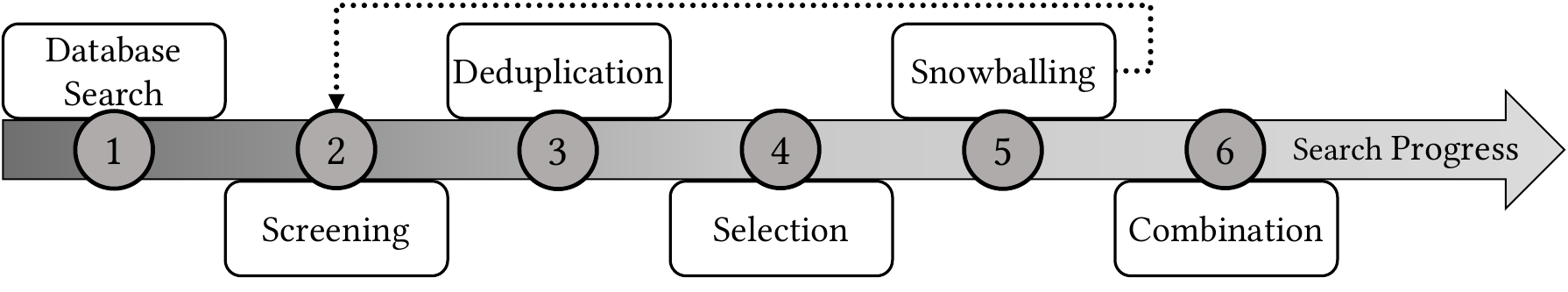}
    \caption{Steps of the literature search.}
    \label{fig:search_process}
\end{figure}

In this section, we elaborate on the search strategy comprising six steps as shown in~\Cref{fig:search_process}.
In Step~1, a database search was conducted across different publication databases to obtain the initial set of potentially relevant publications.
These publications were then screened (Step~2) to prune the entries from the set, which are irrelevant to our research.
After screening, the remaining publications were merged and deduplicated (Step~3) before being checked against precise selection criteria (Step~4) to identify relevant publications for our study.
Additionally, we used the resulting set of relevant publications as a basis for snowballing (Step~5) in order to saturate it with additional relevant publications that were not found during the database search.
Finally, in Step~6, all relevant publications were combined into a single set for the following data extraction and analysis.

\subsubsection{Database Search.}
As highlighted in~\Cref{sec:background}, a number of terms may indicate that a warm-starting technique is employed.
The following list of terms was assembled after screening a set of relevant publications that the authors have identified prior to the start of the study: %
\begin{inparaenum}[(1)]
\item \textit{warm-start},
\item \textit{initial parameter},
\item \textit{initialization strategy},
\item \textit{parameter initialization},
\item \textit{pre-computation},
\item \textit{pre-optimization},
\item \textit{transfer of parameter},
\item \textit{pre-train}, and
\item \textit{transfer learning}.
\end{inparaenum}
These search terms were subsequently used to query electronic publication databases.
For this purpose, a search string was built by concatenating the above terms and linking them to the mandatory term \textit{"quantum"} to find articles related to the quantum computing domain and any of the aforementioned terms.
Thus, the queries are of the general form \[\verb|"quantum" AND ("search_term1" OR "search_term2" OR ...)|.\]
The database search was conducted in early July 2022 and therefore considers publications that appeared by the end of June 2022.
More recent publications were retrieved and included during the subsequent snowballing phase~(see \Cref{subsec:snowballing}).
Each database may apply stemming techniques to automatically consider variations of the search terms, e.g., "pre-compute" instead of "pre-computation".
Hence, one threat to the validity of our study stems from the fact that electronic databases execute searches differently.
Therefore, we tried to mitigate these effects by manually adapting the search queries, including manual stemming.
For the database search, we chose six major electronic publication databases, that appear in guidelines for SMSs in software engineering~\cite{petersen2008guidelines,petersen2015guidelinesUpdate} and are used in such studies~\cite{yussupov2019smsFaaS,varela2021smsSmartContracts,zhao2021smsNLP}, namely: (i) the ACM Digital Library, (ii) IEEE Xplore, (iii) ScienceDirect, (iv) SpringerLink, (v) Wiley Online Library, and (vi) arXiv.
Although SMSs usually focus on formally published works, we decided to include the pre-print database \textit{arXiv} in order to include the most recent progress on the topic.
Many research papers in quantum computing are made available as pre-prints via arXiv before formal publication.
As differentiating between pre-prints and formal publications can provide valuable insights into the current research trends, we treat pre-prints as a separate category in the study.
\looseness=-1

Where possible, the search was limited to the title and abstract of the publications to avoid too many false positives.
We argue that it is reasonable to expect publications with substantial relevancy having either of the search terms in the title or abstract.
However, SpringerLink does not allow restricting the search to the title and abstract of publications.
Therefore, we modified the search query for the full-text search by replacing the \verb|AND| operator with the \verb|NEAR| operator to reduce the number of false positives.
The \verb|NEAR| operator ensures that either of the search terms appear within a distance of ten words, regardless of the order.
Since quantum computing is a highly interdisciplinary field, relevant research articles may also be published in other domains, such as physics and chemistry.
Thus, we did not restrict the search to publications in the computer science field, although most databases support such restrictions.
The exact queries and additional details on the execution of the search for each database are provided in the data repository of this study~\cite{datarepo}.

\subsubsection{Screening.}
\label{strategy_screening}
In the first step, the publications obtained from the database search were screened based on their title and abstract to drop unrelated publications.
In cases where it could not be determined from the titles or abstracts, adaptive reading depth~\cite{petersen2008guidelines} was employed to decide on the publications' relevance.
At this stage, no precise criteria were applied yet, but rather articles that are not related to the topic were excluded, e.g., articles from the quantum physics and quantum chemistry domain that do not describe quantum algorithms but quantum mechanical phenomena.

\subsubsection{Deduplication.}
In the next step, the remaining publications from the different data sources were merged and the identified duplicates are removed.
Duplicates were identified based on the title, author names, and publication year.
In cases where papers had been published as pre-prints before being published as a conference paper or journal article, preference was given to the latter, i.e., to the formally published work.

\subsubsection{Selection.}
\label{subsubsec:strategy_selection_criteria}
After merging and deduplication, the selected publications were filtered based on the precise selection criteria listed below.
Adaptive reading depth~\cite{petersen2008guidelines} was applied where it was sufficient to determine the relevancy of the publication.
The filtering was performed separately and in parallel by $4$ of the authors.
For each publication, the reviewers provided a short rationale why it should be included or excluded.
Borderline cases in which the assessments differed were marked and further discussed until a consensus was reached.
The selection criteria are grouped into inclusion (\cmark) and exclusion (\xmark) criteria:
\begin{itemize}
    \item[\cmark] Publications that introduce warm-starting techniques in the context of quantum algorithms.
    \item[\cmark] Publications that apply or evaluate such warm-starting techniques.
    \item[\cmark] Publications that facilitate the application of warm-starting in the context of quantum algorithms.
    \item[\xmark] Publications that are not written in English.
    \item[\xmark] Secondary or tertiary studies lacking an own contribution to the topic itself.
    \item[\xmark] Publications that are not available as full-text or not in the form of a full research paper, e.g., (extended) abstracts, presentations, tutorials, research proposals, demo papers, etc.
    \item[\xmark] Publications that do not provide sufficient insight on details, that are required for the analysis of the methods.
\end{itemize}

\subsubsection{Snowballing.}
\begin{figure}[h]
    \centering
    \includegraphics[page=3,width=0.55\linewidth,trim=0 0 485 205,clip]{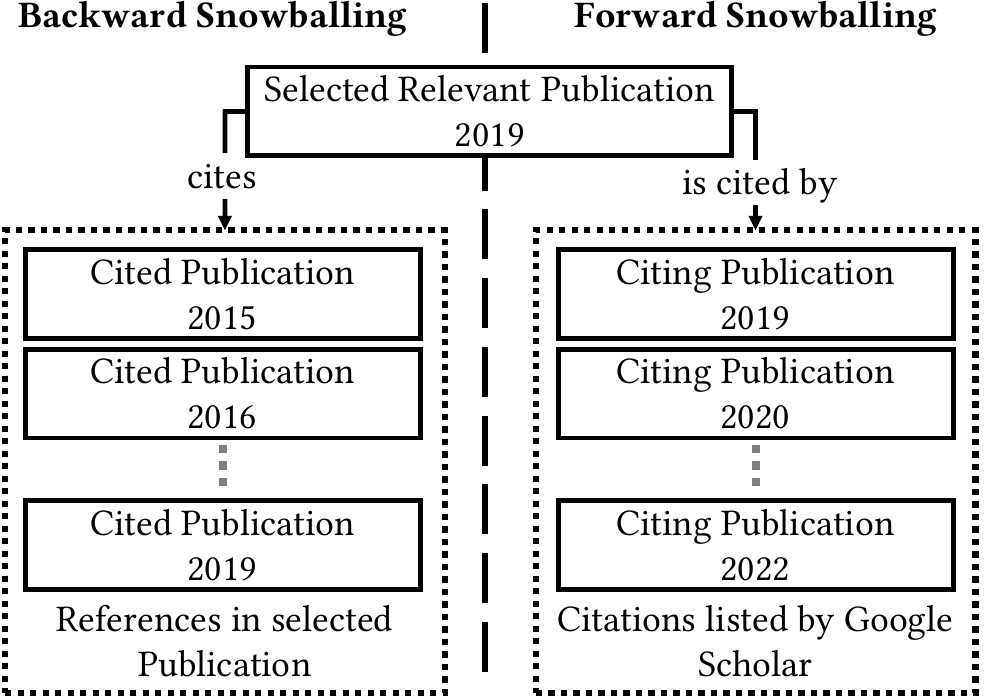}
    \caption{Scopes of backward and forward snowballing.}
    \label{fig:snowballing}
\end{figure}

Since we can not assume that all relevant publications were found during the database search, we applied the snowballing technique~\cite{jalali2012snowballing,wohlin2014snowballing}, particularly forward and backward snowballing.
As shown in \Cref{fig:snowballing}, the search scope for backward snowballing are the references of the selected publications, whereas in forward snowballing, we likewise consider papers that cite selected publications.
For backward snowballing, we directly checked the references listed in a publication, whereas for forward snowballing, Google Scholar was used as a tool to find its citations.
These references and citations were processed analogous to Steps~2--4 outlined above (see \Cref{fig:search_process}).
As the snowballing was conducted in late September 2022, our study also considers relevant publications that appeared after the database search was carried out.
\looseness=-1
\label{subsec:snowballing}

\subsubsection{Combination.}
\label{subsec:combination}
In the final step, the set of relevant publications was consolidated by combining the results obtained from the database search and snowballing into a single dataset.
Furthermore, some publications may be related to the same concept and research, e.g., a journal extension of a previously published conference paper.
We analyzed such relations, linked these papers, and treated them as a single publication entity describing the same concept.

\subsection{Data Extraction and Analysis}
\label{subsec:data-extraction}
After completion of the search phase, we extracted and analyzed relevant data from the final set of publications to answer our research questions.
We used the bibliography management tool \textit{JabRef}\footnote{\url{https://github.com/JabRef/jabref}} for curating the final set of publications and online spreadsheets for the curation of extracted data, as they facilitate the collaboration among the researchers.
Using spreadsheets also helps avoid manual errors since the aggregation of data, such as conditional counting, is automated.
To conduct the data extraction among four researchers, the set of publications was split into chunks for each researcher to extract the relevant data from.
To sustain consistency, the data extraction was reviewed by the first author and discussed among the researchers where necessary to achieve a consensus.
\Cref{tab:data_extraction} provides an overview of the data extracted from the publications to answer each research question.
For each entry, the respective range of values or exemplary values are provided.

\begin{table}[b]
    \centering
    \caption{Overview of data to be extracted from the publications for each research question. (WS stands for warm-starting, for brevity.)}

    \begin{tabularx}{\linewidth}{lp{4.15cm}X}
        \toprule
        \textbf{RQ (Focus)} & \textbf{Data to Be Extracted} & \textbf{Value Range/Example Values} \\ \midrule
         
        \multirow{6}{*}{\makecell[l]{RQ1\\(publication\\trends)}} 
        & \textbullet\ Publication date & e.g., June 2020 \\
        & \textbullet\ Type of publication venue & $\in\{$Journal, Conference, Workshop, Pre-Print$\}$ \\
        & \textbullet\ Publication venue & e.g., IOP Quantum Science and Technology (Journal)\\
        & \textbullet\ Author affiliations & $\in\{$academic, other research institutions, industry, mixed$\}$ \\ 
        & \textbullet\ Type of research & $\in$\ \{Validation Research, Evaluation Research, Solution Proposal, Philosophical Paper, Opinion Paper, Experience Paper$\}$~\cite{petersen2008guidelines,wieringa2006requirements}
        \\  \hline
        
        \multirow{3}{*}{\makecell[l]{RQ2\\(algorithms \&\\benefits)}} 
        & \textbullet\ Quantum algorithm(s) & e.g., QAOA, VQE, QNN \\
        & \textbullet\ Computational problem(s) & e.g., Maximum Cut, Portfolio Optimization, Classification\\ 
        & \textbullet\ Stated goals or benefits of WS & e.g., speedup, accuracy increase, reduction of resource consumption \\
         \hline
        
        \multirow{2}{*}{\makecell[l]{RQ3\\(techniques)}} 
        & \textbullet\ Keywords & e.g., quantum circuit, annealer, parameter initialization, initial state, classical-to-quantum, machine learning \\
        \bottomrule
    \end{tabularx}
    \label{tab:data_extraction}
\end{table}

To answer RQ1, which focuses on publication trends, we extracted bibliographic information, such as the publication year and month, publication venue, and author affiliations. 
Additionally, to categorize the publications by the type of research they employ, we used an established classification scheme for research types~\cite{petersen2008guidelines,wieringa2006requirements}.
Particularly, empirical research on the topic, such as validation research and evaluation research, and non-empirical research, including solution proposals and experience papers, had to be classified.
While RQ1 focuses on the publications as items of interest, the focus of RQ2 and RQ3 is on the warm-starting techniques themselves.
Since a single relevant publication can cover multiple techniques, the data extraction for RQ2 and RQ3 was conducted per technique, i.e., when more than one distinct technique was encountered in a publication, we duplicated the entry in the spreadsheet and subsequently extracted the data for each technique separately.
The first part of RQ2 focuses on computational problems and quantum algorithms involved in a warm-start.
Therefore, we extracted the names of the quantum algorithms involved in each technique and the computational problems tackled.
Additionally, to answer the second part of RQ2, we documented the stated goals and benefits of warm-starting techniques in the quantum domain by taking notes of the improvements the researchers aimed for and managed to achieve, e.g., a speedup or increase of accuracy.
Using the keywording technique~\cite{petersen2008guidelines,petersen2015guidelinesUpdate}, we further grouped the noted core drivers and improvements into a number of relevant aspects.
As the classification of warm-starting techniques aimed at with RQ3 first requires further analysis of the techniques, we started by extracting keywords~\cite{petersen2008guidelines,petersen2015guidelinesUpdate} for each technique.
Such keywords resemble potentially relevant properties of a technique that can be used for the classification.
Examples are listed in \Cref{tab:data_extraction} and include information on the quantum devices used, and parts of the algorithms affected by the warm-start, etc.
The keywords extracted for RQ3 were then used to compile a list of relevant properties of warm-starting techniques, that can be utilized for their eventual classification~(see~\Cref{sec:results}).
To achieve this, the publications were examined and discussed by at least $4$ of the authors to reach consensus on the relevant properties and a classification scheme based thereon that was subsequently discussed with all co-authors for further refinement.
\looseness=-1

%% file: content/Results.tex

\section{Results and Discussion}
\label{sec:results}
In this section, we present and discuss the results of our study.
First, we elaborate on the results from the literature search process.
Afterward, we focus on the results obtained through the data extraction described in~\Cref{subsec:data-extraction}.
In this regard, we present and discuss the results of the analysis of the relevant publications, starting with the publication trends followed by the analysis of quantum algorithms involved in warm-starts and the goals and benefits of warm-starting as stated in the analyzed research literature.
Finally, we present a list of relevant properties of warm-starting techniques and introduce our classification scheme to categorize them.

\subsection{Search Results}
\begin{figure}[h]
    \centering
    \includegraphics[page=1,width=0.80\linewidth,trim=0 230 270 0,clip]{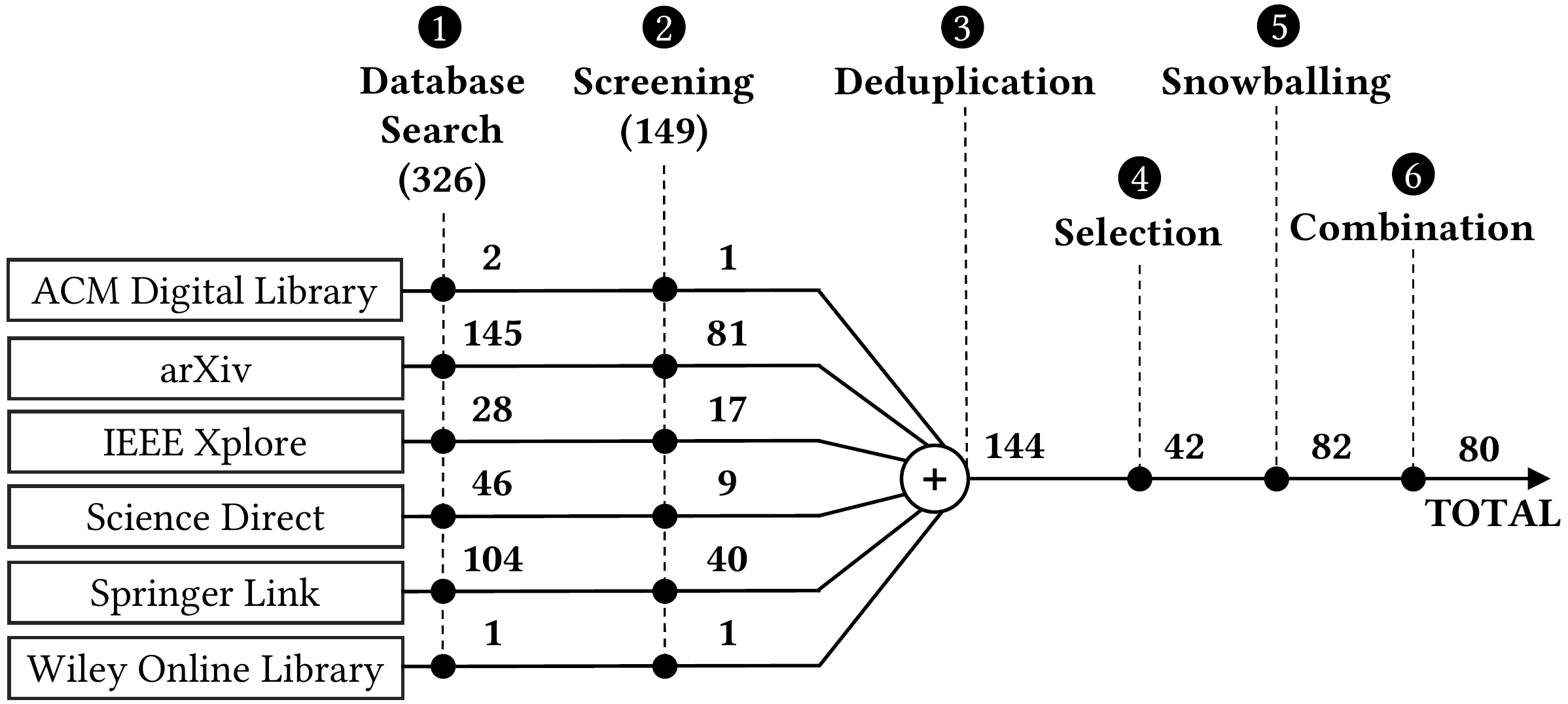}
    \caption{Results of the multiphase search and selection process (figure based on~\cite{yussupov2019smsFaaS,di2017smsArchitectureMicroservices}).}
    \label{fig:search_results}
\end{figure}

The number of search results obtained from each database and after each of the steps outlined in~\Cref{subsec:search_strategy} is presented in~\Cref{fig:search_results}.
The database search yielded a total of $326$ publications across all databases.
After the initial screening, a total of $149$ publications remained for an in-depth analysis against the selection criteria.
This number was reduced to $144$ publications by merging and deduplicating the results from different databases.
Applying the selection criteria listed in~\Cref{subsubsec:strategy_selection_criteria}, we found $42$~publications to be relevant to the topic.
An additional $40$ relevant publications were found via backward and forward snowballing.
In the combination step, two cases were identified in which two publications, respectively, described nearly identical work.
In both cases, these two entries were combined to be treated as one entity in our study.
The final set of $80$ publications selected for further analysis is listed in~\Cref{sec:appendix-publications}.
We assigned unique identifiers to the publications (P01-P80) by which we will refer to them in the remainder of this section.

\subsection{Publication Trends}
In this section, we present the results with respect to the publication trends to answer RQ1.

\subsubsection{Number of Publications}
\begin{figure}[t]
    \centering
    \includegraphics[page=5,width=0.65\linewidth,trim=0 295 550 20,clip]{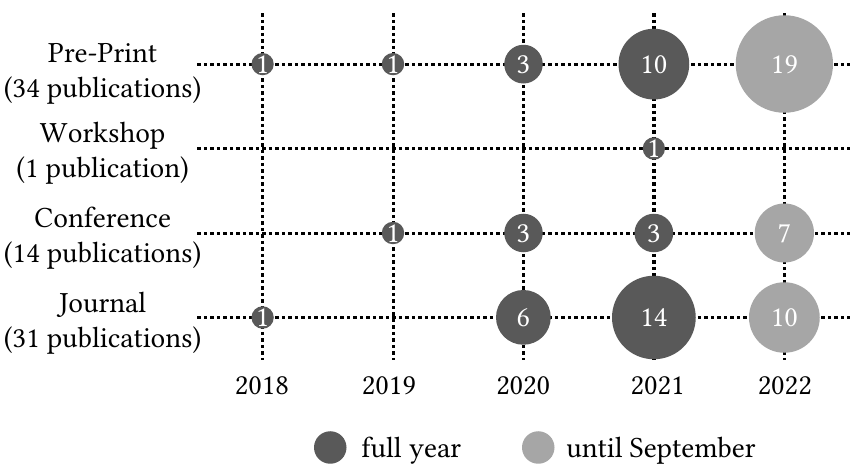}
    \caption{Publications per year and type of publication venue.}
    \label{fig:results-publication-venues}
\end{figure}

As shown in \Cref{fig:results-publication-venues}, the overall number of publications relevant to the topic increased over the course of the last few years.
Two first publications appeared in late 2018.
However, the number of relevant publications started increasing significantly only two years later, when it increased to 12 in 2020.
In the following two years, it grew to 28 publications in 2021 and, until the conduction of this study, to 36 publications in 2022.

\definecolor{dark_gr}{RGB}{89, 89, 89}
\definecolor{light_gr}{RGB}{166, 166, 166}

\subsubsection{Publication Venues}
\Cref{fig:results-publication-venues} further shows the publication trends w.r.t. the types of publication venues.
Peer-reviewed publications are predominantly presented in journals and conferences, whereas only one workshop paper is present in our final dataset.
A large portion of relevant publications are published as pre-prints.
The trend of growing publication numbers can also be observed for these individual types of venues, except for workshops.

The most popular venues, besides arXiv for pre-prints (34~publications), were three journals and one conference (3~publications each): ``IOP Quantum Science and Technology'', ``Quantum'', ``Springer Quantum Machine Intelligence'', and the ``IEEE International Conference on Quantum Computing and Engineering (QCE)''. 
The next place is shared among the ``World Scientific International Journal of Quantum Information'', ``Physical Review Research'', and the ``IEEE International Conference on Innovative Trends in Information Technology (ICITIIT)'' (2 publications each).
The remaining 28 publications are distributed across venues that appeared only once in the dataset.

\subsubsection{Author Affiliations}
\begin{figure}
    \centering
    \begin{subfigure}[b]{0.53\textwidth}
        \centering
        \includegraphics[page=6,width=\linewidth,trim=0 325 570 15,clip]{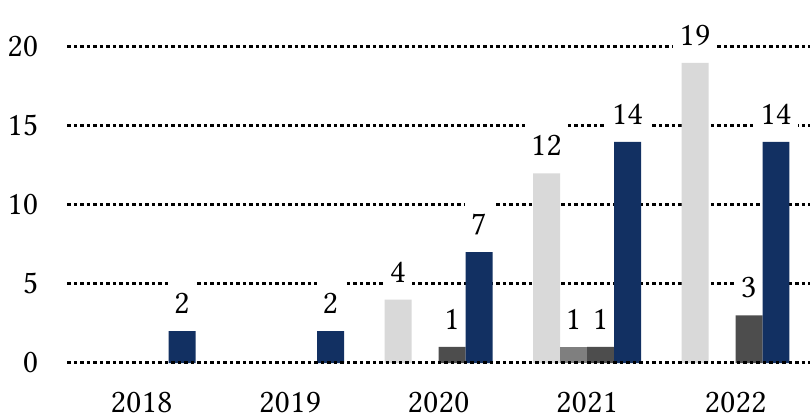}
        \caption{Publications per year by author affiliations.}
        \label{fig:results-affiliations-subfig1}
    \end{subfigure}
    \hfill
    \begin{subfigure}[b]{0.465\textwidth}
        \centering
        \includegraphics[page=7,width=\linewidth,trim=0 375 625 0,clip]{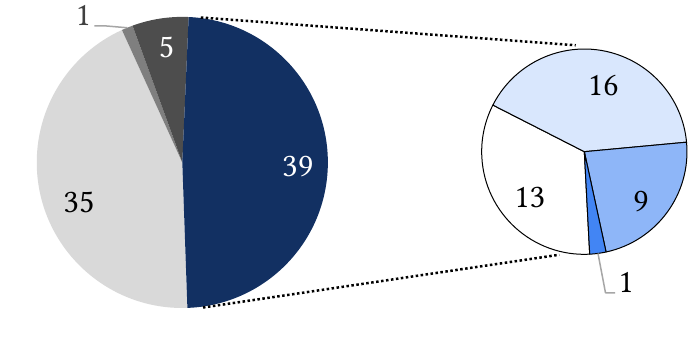}
        \caption{Overall distribution of author affiliations (left) and detailed mixed author affiliations (right).}
        \label{fig:results-affiliations-mixed}
    \end{subfigure}
    
    \hspace{12mm}
    \begin{subfigure}[b]{0.7\textwidth}
        \includegraphics[page=7,width=\linewidth,trim=0 280 435 190,clip]{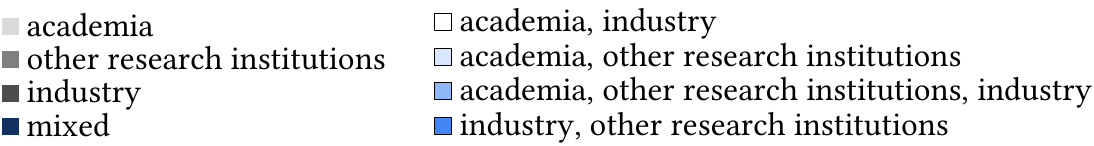}
    \end{subfigure}
        \caption{Author affiliations of relevant publications.}
        \label{fig:results-affiliations}    
\end{figure}
The author affiliations observed in the dataset are presented in \Cref{fig:results-affiliations}.
\Cref{fig:results-affiliations-subfig1} shows a continuous growth of research by authors affiliated with academic institutions, such as technical and research universities.
A similar growth was observed for mixed author affiliations, i.e., involving authors with different types of affiliations.
Research conducted solely by authors from industry or research institutions outside academia is comparatively rare.
However, the number of publications from the industry sector started increasing recently, possibly indicating a growing interest of the industry to employ warm-starting to enable practically useful quantum applications.\looseness=-1

\Cref{fig:results-affiliations-mixed} summarizes the distribution of author affiliations in the overall dataset and provides a more detailed view on mixed affiliations.
As illustrated by the overall numbers, mixed affiliations are the most frequent among the relevant publications.
The vast majority of mixed affiliations includes authors from academia.
Thus, a total of 73 publications ($\approx 91\%$) were at least co-written by authors affiliated with academic institutions, whereas only a total of 28 publications and 27 publications were (co-)written by industry and other research institutions, respectively.

\subsubsection{Research Types}
Nearly all relevant publications fall into the category of validation research in the research classification scheme employed here~\cite{petersen2008guidelines,wieringa2006requirements}.
The only exception is one publication (P05) that was categorized as a solution proposal since it provides only a relatively small proof-of-concept example for the technique proposed therein.
A lack of evaluation research is plausible since evaluation research refers to the evaluation of a technique that is used in practice.
However, quantum algorithms are still far from being usable in practice due to the limitations of NISQ devices.
Similarly, experience papers refer to experience collected in practice and could therefore not be anticipated in the dataset.
Moreover, we have not observed philosophical papers or opinion papers.
Therefore, validation research and solution proposals are the only types of research observed in the dataset.
The fact that validation research is dominant in the dataset shows a clear trend that techniques are not only proposed, discussed, or opinionated about, but typically tried out on sample data to illustrate their benefits numerically.

\subsection{Problems, Quantum Algorithms, and Benefits}
\label{subsec:problems_algos_benefits}
To answer RQ2, we analyzed the extracted data w.r.t. the problems tackled with the warm-started algorithms, the quantum algorithms involved in the warm-starts, and the stated goals and benefits of the use of warm-starting.
The 80 publications in the dataset discuss a total of 88 techniques relevant to our study, i.e., there are a few papers that propose, discuss, or validate more than one technique.
We assigned identifiers to each technique that follow the pattern of P08-T1 referring to the first technique and P08-T2 to the second technique observed in publication P08, respectively.
Techniques from publications that contain only one relevant approach are referred to by the publication identifier for brevity, e.g., P01 instead of P01-T1.
The results are presented and discussed in the following three subsections, each covering one of the aforementioned aspects.
\looseness=-1

\subsubsection{Computational Problems and Quantum Algorithms in Warm-Starts}
\input{figures/table_computational_problems.tex}
\input{figures/table_quantum_algos.tex}
First, we analyzed the extracted data regarding the problems solved with warm-started algorithms and the quantum algorithms involved in the warm-starts.
\Cref{tab:results-problems,tab:results-algos} summarize the results we discuss in this section by relating the computational problems and quantum algorithms with each other.
In~\Cref{tab:results-problems}, we provide a list of computational problems tackled with warm-started algorithms, their number of occurrences and the identifiers of related techniques.
One warm-starting technique can be used in the context of multiple problems, thus, the total number of occurrences of problems presented in~\Cref{tab:results-problems} is larger than the number of techniques and amounts to 121.
Analogously, one warm-starting technique can be applicable to or involve multiple, sometimes overlapping (categories of) quantum algorithms, e.g., P03 mentions both QAOA and VQE as targeted algorithms, but also PQC in general, and P80 is likewise applicable to both QAOA and VQE.
Therefore, the total number of occurrences of algorithms presented in~\Cref{tab:results-algos} is also larger than the number of techniques.
We found that a total of 16 quantum algorithms are involved 103 times in warm-starting techniques.
However, the information is provided on different abstraction levels across publications, which limits the quality of the extracted information.
Some authors directly indicate that their technique can be applied in the context of QAOA or even specific variants of QAOA, while others mention more general categories and concepts such as VQA and PQC, that can be considered as umbrella terms including QAOA.
In~\Cref{tab:results-algos}, we provide a list of quantum algorithms (or categories) as they were stated in the publications, their number of occurrences as well as the identifiers of related techniques.\looseness=-1

As shown in~\Cref{tab:results-problems}, the most frequently tackled problems are classification problems, which occurred 33 times in the dataset.
The subgroup of image classification problems also occupies the third place in~\Cref{tab:results-problems}.
Note, that there is a large overlap between techniques that are applied in the context of classification problems and techniques that involve QNN~(cf.~\Cref{tab:results-algos}), reflecting the relevancy of QNN for classification problems.
The second most frequent problem is the Maximum Cut problem (MaxCut), a well-known NP-complete graph partitioning and optimization problem, that is frequently employed as an example problem in quantum optimization.
The large overlap of techniques involving both MaxCut and QAOA in~\Cref{tab:results-algos} and~\Cref{tab:results-problems}, respectively, confirms the popularity of MaxCut as a demonstration problem for QAOA in particular.
Other frequent problems occurring in the dataset include Ground State Estimation and Ground State Preparation problems, which appear primarily in conjunction with QAOA, VQE, and the Hamiltonian Variational Ansatz.
Traverse Field Ising Models are likewise tackled with QAOA and the Hamiltonian Variational Ansatz.
Due to the aforementioned overlaps, the QAOA, QNN, and VQE are also the three most frequent algorithms in~\Cref{tab:results-algos}.
This clearly shows that the focus of warm-starting techniques is on variational algorithms requiring the optimization of parameters in quantum circuits.
This trend continues in the next entries of~\Cref{tab:results-algos} with specific ansätze for VQAs, such as the Hamiltonian Variational Ansatz and Hardware-Efficient Ansatz, and more general concepts of variational algorithms, such as Variational Quantum Circuits and PQC.
Only a few exceptions focus on quantum annealing and adiabatic quantum computing, as well as other specific quantum algorithms such as Quantum Phase Estimation, Quantum Circuit Born Machines, Continuous-Time Quantum Walks, and quantum versions of Boltzmann Machines.

\mrev{Overall, \Cref{tab:results-algos,tab:results-problems} include mostly typical NISQ algorithms and problems tackled with them.
Although the literature search and selection was not specifically restricted to NISQ algorithms, these results highlight the wide usage of and the importance of studying warm-starting techniques in the NISQ era.}

\subsubsection{Goals and Benefits of Warm-starting}
\begin{figure}[b]
    \centering
    \includegraphics[page=9,width=1\linewidth,trim=17 370 365 0,clip]{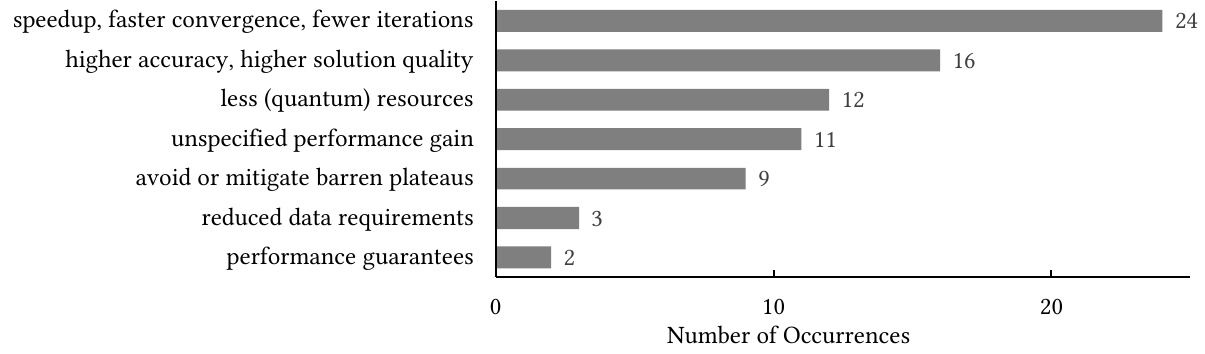}
    \caption{Goals and benefits of warm-starting in the quantum computing domain and their number of occurrences in the dataset.\looseness=-1}
    \label{fig:results-motivation}
\end{figure}
Last, we analyzed the extracted data regarding the goals and benefits of the warm-starting techniques.
Since the textual descriptions found in the publications vary widely from each other, we grouped the goals and benefits into seven aspects, as presented in~\Cref{fig:results-motivation}.
Note that we do not look at the general goals and benefits of quantum algorithms here, but those of the warm-starting techniques specifically.
Some publications did not make specific claims, while others reported goals and benefits falling into multiple categories.
The overall number of occurrences is slightly lower than the number of publications and techniques.
Evidently, the most frequent aspect is focused on speedups which are achieved, for example, through a reduction of iterations of classical optimization loops in VQA (e.g., P25, P31, P63, P77) or otherwise accelerated convergence (e.g., P27, P61).
A second important factor is accuracy or solution quality, which is particularly noteworthy, since a common perception of quantum computing in general is focused on speedups rather than quality improvements.
However, as shown here, some warm-starting techniques in the area, and therefore also quantum algorithms, actually focus on improving the quality of solutions, e.g., by achieving more accurate classifications (e.g., P59, P60) or approximations (e.g., P15, P55-T1, P79).
The previously mentioned speedups can in some cases also mean a reduction of resources needed to run an algorithm, which is reflected in the third most frequent aspect.
Particularly for quantum algorithms in contemporary quantum computing, it is relevant to reduce resource consumption since quantum devices are still scarce and need to be shared among thousands of users.
However, this aspect is likely to remain relevant even beyond the NISQ era, as the general awareness for the reduction of resource and energy consumption is becoming a dominant factor in the society and in this particular case will continue to facilitate the efficient usage of quantum computing resources.
``Unspecified performance gain'' in~\Cref{fig:results-motivation} refers to mentions of a performance improvement that is not specified clearly enough to be attributed to one of the other aspects, such as speedup or accuracy increase.
Barren plateaus are areas in the parameter space of cost functions of optimization problems
with vanishing gradients and thus hinder the optimization process significantly.
Therefore, avoiding barren plateaus is a key factor for improving the optimization process of VQAs.
Obviously, this aspect is also related to the aforementioned speedups, accuracy improvements, and reduction of resource consumption, since avoiding barren plateaus goes hand in hand with these achievements.
Another less frequently mentioned aspect is the reduction of data requirements, e.g., in the sense that fewer data points are needed when experience from previous computations is employed (e.g., P30, P38).
Additionally, there were two cases (P04, P41) in which performance guarantees could be obtained or retained through warm-starts, for example, when a classical algorithm with such guarantees is utilized to generate a starting point for warm-starting a quantum algorithm.
For instance, the technique of~\citet{egger2021warm} (P04) utilizes the classical Goemans-Williamson Algorithm (GW)~\cite{goemans1995improvedapproximation} for MaxCut to generate an initial state for warm-starting QAOA and thus retains the approximation ratio achieved by GW in the warm-started QAOA algorithm.
\looseness=-1

\subsection{Warm-Starting Techniques}
\label{subsec:results-ws-techniques}
This section presents the results for RQ3, i.e., it focuses on warm-starting techniques and their classification.
Following the research design outlined in~\Cref{sec:researchmethod}, we first compiled a list of relevant properties of warm-starting techniques and subsequently a classification scheme for these techniques based on their properties.
The following subsections describe the properties and classification scheme in more detail.

\subsubsection{Properties of Warm-Starting Techniques}
\input{figures/table_properties.tex}
Based on the extracted keywords, we compiled the list of properties presented in~\Cref{tab:results-properties}.
We distinguish seven relevant properties of warm-starting techniques in the quantum computing domain:
The first property shown in~\Cref{tab:results-properties} captures the \textit{Direction} in which a technique is applied, i.e., which types of algorithms, classical or quantum, take the role of the source and target algorithm, respectively, in the warm-starting procedure.
For example, the direction is classical-to-quantum (C2Q) when a classical procedure generates the information utilized to warm-start a quantum algorithm (e.g., P04, P09-T1, P15).
The next property is the kind of \textit{Quantum Devices} involved, namely quantum annealers, gate-based quantum computers, or, possibly, both in the Q2Q warm-starting case (P13).
Further, we document different kinds of \textit{Entry Points} for the warm-starting techniques in the eponymous property as shown in~\Cref{tab:results-properties}, i.e., we identify where the warm-start is applied in the targeted algorithm.
For example, there exist warm-starting techniques where values obtained from the source algorithm are encoded into the initial state of the quantum circuit of the target algorithm (e.g., P04, P15), as opposed to examples where the source algorithm is employed to obtain initial parameters for the variational optimization of the target algorithm (e.g., P05, P13, P47).
Thus, in the former case, the entry point of the warm-starting technique is the encoding in the initial quantum state, whereas in the latter case, the initialization of the variational parameters resembles the entry point.
Moreover, it is important to understand whether a quantum algorithm involved in warm-starting needs to be adapted to be compatible with the chosen technique (e.g., P04, P15, P75).
\Cref{tab:results-properties} captures this requirement as the \textit{Change of Quantum Algorithm} property.
We also discovered that some techniques can be reflexive, in the sense that the target algorithm is also used as the source of information for the warm-start. 
For instance, if a variational algorithm is executed on small problem instances in order to obtain initial parameters for a larger targeted problem instance, the underlying warm-starting technique is categorized as reflexive (e.g., P05, P21, P70).
On the other hand, ``non-reflexive'' warm-starts are based solely on information obtained from a different algorithm or procedure (e.g., P04, P09-T1, P13).
In~\Cref{tab:results-properties} this is captured as the \textit{Reflexivity} property.
\textit{Data Generation} is another property of warm-starting techniques: We observed that some techniques require solely the targeted problem instance in order to apply a warm-start (e.g., P04, P13, P15), whereas other techniques require data gathered from the execution of additional, possibly related, problem instances (e.g., P06, P07, P09-T1).
Lastly, there are cases where such data is processed using Machine Learning (ML) or Quantum Machine Learning (QML) techniques, while others require trivial or no \textit{Intermediary Processing} (summarized as ``None'', e.g., P04, P13, P15).
For example, an ML model (e.g., P06, P07, P09-T1) or QML model (P09-T2, P09-T3, P50) is trained based on gathered data.
\looseness=-1

These properties were evaluated for the dataset, such that each technique takes exactly one of the values described in~\Cref{tab:results-properties} w.r.t. each property.
The number of techniques exhibiting a certain value is provided in the right-most column of the table.
The list of properties can be further extended with additional relevant properties or values in the future, e.g., there may be more entry points for warm-starts that have not occurred in our dataset.
For the next step, the techniques were grouped according to the combinations of these seven properties.
We thus obtained a total of 25 groups of techniques, with up to 26 techniques per group.

\subsubsection{Classification Scheme for Warm-Starting Techniques}
\begin{figure}
    \centering
    \includegraphics[page=8,width=0.95\linewidth,trim=0 102 540 0,clip]{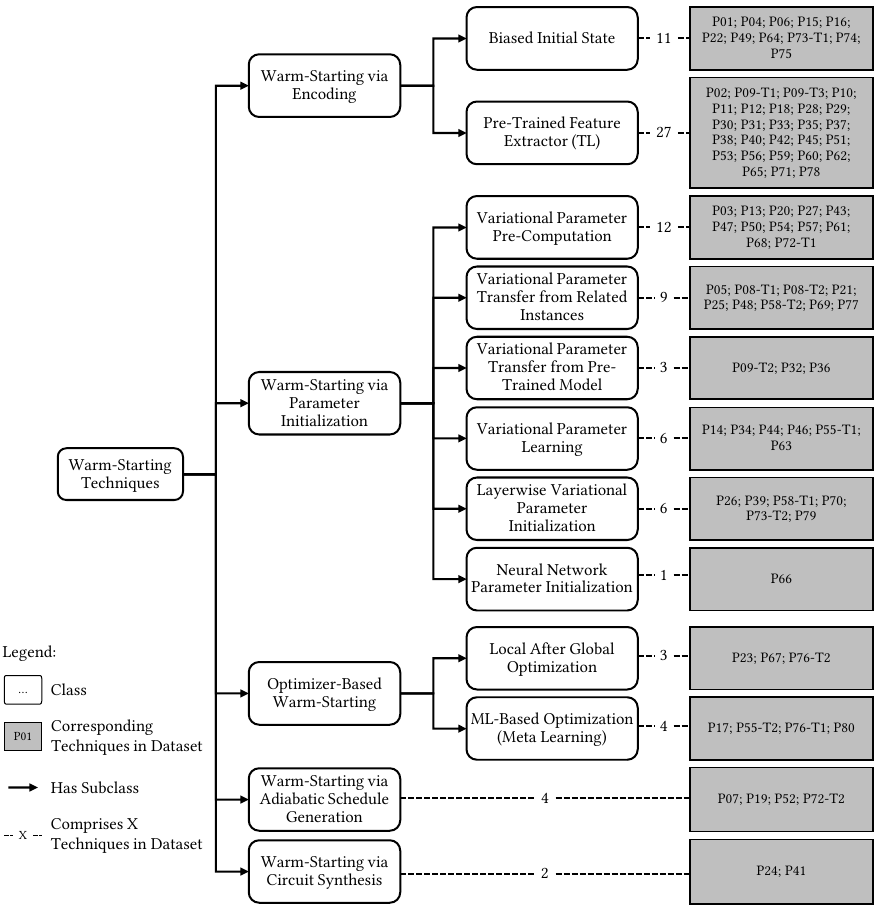}
    \caption{Classification scheme for warm-starting techniques and identifiers of the techniques in the dataset belonging to each class.}
    \label{fig:results-classification-scheme}
\end{figure}

In the final step, we derived the classification scheme for warm-starting techniques by analyzing the groups of techniques that show similar combinations of values w.r.t. the properties identified in the previous step.
The scheme is presented in~\Cref{fig:results-classification-scheme} comprising five major classes of warm-starting techniques.
For the most populated major classes, the classification scheme provides a total of ten refined subclasses.
Additionally, \Cref{fig:results-classification-scheme} lists the number of techniques per class and their identifiers in the dataset.
The major classes are derived from the entry points, which we identified as the most characteristic property suitable to distinguish different kinds of warm-starting techniques.
The encoding-related and parameter-related entry points for the classical and quantum domains were grouped together in the \textit{Encoding} and \textit{Parameter Initialization} classes, respectively.
In the following, we characterize each of the classes shown in~\Cref{fig:results-classification-scheme} by elaborating on the warm-starting mechanisms and the properties of the techniques contained in them.

\smallskip\noindent
\textbf{Warm-Starting via Encoding.} This first major class shown in~\Cref{fig:results-classification-scheme} focuses on techniques that warm-start algorithms through the encoding of information into the initial state of quantum or classical algorithms.

\begin{itemize}
\item \textit{Biased Initial State}.
Techniques in this subclass are focused on gate-based quantum devices and occur in the C2Q or Q2Q warm-starting direction~(see~\Cref{tab:results-properties}). 
The entry point in this class of techniques is always the encoding in the initial quantum state, since these techniques aim to bias the initial state towards favorable solutions such that the algorithm's ability to find a suitable solution or approximation is improved.
A prominent example is the warm-started QAOA for MaxCut that utilizes the classical GW algorithm~(P04).
It is typical for this class of techniques that the quantum algorithm is adapted for the encoding. 
Moreover, most techniques can be applied solely based on the targeted problem instance, i.e., they merely pre-process the target instance. 
However, we also found a technique in this class, that relies on training a classical model for a family of problem instances (P06).
\looseness=-1

\item \textit{Pre-Trained Feature Extractor (TL)}. 
The second subclass of techniques in the \textit{Encoding} class was also found to be the most frequent in the dataset.
With one exception, these techniques are C2Q warm-starts for gate-based quantum computers, which rely on a classical ML model trained on data other than the targeted problem instance to encode meaningful information (features) in the initial quantum state.
The exception is a Q2C technique (P09-T3), that relies on a QML model trained on data other than the targeted problem instance and passes features into a classical neural network.
Therefore, these techniques are unanimously non-reflexive.
We named this class after the characteristic that a pre-trained model is utilized to extract informative features from the targeted problem instance.
However, these techniques are also often called ``transfer learning'' (TL) after their counterparts in classical machine learning~\cite{mari2020transfer}.
The idea behind transfer learning is that a model trained on a problem can be trained further to solve a new but similar or related problem, requiring less effort than training a new model from scratch.
C2Q transfer learning schemes make use of a classical ML model that is already trained for a task by transforming it into a feature extractor.
For example, a trained classical neural network for a general image classification task can be pruned of its final layers to obtain intermediate values that represent informative features of the input.
This intermediate information can then be encoded into the initial quantum state to solve a more precise classification task.
Note that transfer learning in classical computing, especially in the context of neural networks, has also often been referred to as warm-starting~\cite{perrone2018scalable,ash2020warm,gavves2015active,tirumala2022novel}. 
Therefore, considering such techniques in the quantum computing domain as warm-starting techniques is in line with existing research on transfer learning and warm-starting techniques in the domain of classical computing.
\end{itemize}

\smallskip\noindent
\textbf{Warm-Starting via Parameter Initialization.} This major class of techniques shown in~\Cref{fig:results-classification-scheme} focuses on the improved initialization of parameters.
\vspace{-0.75mm}
\begin{itemize}
\item \textit{Variational Parameter Pre-Computation.}
These techniques aim to pre-compute parameters for variational quantum algorithms. 
Therefore, these are mostly C2Q warm-starts, but also Q2Q techniques exist in this group (P13, P50, P54).
The entry point is always the parameter initialization, where pre-computed parameters are plugged in to achieve an advantage.
Most techniques in this group do not require adaptations of the quantum algorithm or intermediary processing, however, in one case (P68) an adaptation is needed.
Some techniques are reflexive in the sense that they rely on classical simulation of simplified quantum circuits (P27, P68).
\looseness=-1

\item \textit{Variational Parameter Transfer from Related Instances.}
As indicated by the name, techniques in this class utilize known (near-)optimal parameters from other problem instances to initialize variational parameters for the targeted problem instance.
Therefore, these Q2Q techniques are reflexive and require data from other problem instances.
However, there are typically no changes of the quantum algorithm required and no intermediary processing in the form of (Q)ML involved.
Clearly, these techniques make use of the intuition that optimal parameters for one problem instance may be located close to optimal parameters of similar problem instances.
Therefore, the optimization may be started in the vicinity of the actual optimum and thus require either fewer iterations to achieve the same solution quality as the cold-started algorithm or can obtain better solutions with the same optimization effort.

\item \textit{Variational Parameter Transfer from Pre-Trained Model.}
Although related to the previous class, these Q2Q techniques rely on a pre-trained QML model whose variational parameters are utilized as initial parameters for a new model.
Hence, they apply Q2Q transfer learning~\cite{mari2020transfer}.
These techniques also do not require changes of the quantum algorithm but utilize additional data from other problem instances to pre-train the source model.
In contrast to the previous class, they can be non-reflexive, e.g., when parameters are transferred from a smaller pre-trained model to an extended target model.

\item \textit{Variational Parameter Learning.}
This subclass is related to the previous two.
However, it relies on training ML models to initialize variational parameters in contrast to direct parameter transfers, i.e., the models aim to learn from previous executions of the algorithm for different problem instances to generate good initial parameters for a new problem instance.
Therefore, the techniques in this class are reflexive Q2Q techniques and require data from the execution of the quantum algorithm for other problem instances.

\item \textit{Layerwise Variational Parameter Initialization.}
This class of techniques applies an optimization procedure known as layerwise learning~\cite{skolik2021layerwise}.
Variational parameters are not optimized all at once, but rather in groups of few parameters.
Once a group of parameters has been optimized, the optimal values are transferred as initial parameters to another incarnation of the variational parameter optimization for the next group of parameters.
In contrast to previously discussed techniques, the transfer of parameters in this class happens between different runs of the same quantum algorithm for the same problem instance.
Therefore, these reflexive techniques are characterized by the Q2Q direction and require only the targeted problem instance without changing the quantum algorithm or applying intermediary processing.\looseness=-1

\item \textit{Neural Network Parameter Initialization.}
This last subclass of parameter initialization techniques accounts for techniques that initialize neural network parameters, in particular, in a Q2C setup.
In contrast to the parameter initialization techniques described thus far, the entry point in this class is the parameter initialization in a \textit{classical} neural network.
Although we have only one technique assigned to this class (P66), it seems safe to assume that such techniques are generally non-reflexive due to their Q2C nature and do not require changing a quantum algorithm since the entry point is on the classical side.

\end{itemize}

\smallskip\noindent
\textbf{Optimizer-Based Warm-Starting.} This major class of techniques depicted in~\Cref{fig:results-classification-scheme} focuses on improving the optimization process.
In contrast to previous techniques, these techniques not only concern the pre-computation or transfer of initial parameters but adapt the optimizer or optimization procedure as a whole.

\begin{itemize}
\item \textit{Local After Global Optimization.}
As the name indicates, these techniques rely on plugging together multiple optimization strategies and utilizing optimized parameters from one strategy to initialize, i.e., to warm-start, the subsequent optimization.
Often, these procedures combine a global optimization strategy to find an area of interest in the parameter landscape that is then further explored by a local optimization strategy.
Therefore, these Q2Q techniques are reflexive and do not require changes of the quantum circuit or data from additional problem instances.

\item \textit{ML-Based Optimization (Meta Learning).}
The second subclass of the \textit{Optimizer} class covers techniques that rely on meta learning.
Their goal is to train an ML model that acts as an optimizer for variational quantum algorithms.
Therefore, these Q2Q techniques utilize data collected from the optimization of a set of additional problem instances to learn an optimization strategy.
This makes these techniques reflexive and relying on ML as an intermediary processing step.
Since they concern the optimization process outside the variational quantum circuits, the quantum algorithms themselves are not changed.

\end{itemize}

\smallskip\noindent
\textbf{Warm-Starting via Adiabatic Schedule Generation.}
This major class of techniques shown in~\Cref{fig:results-classification-scheme} acts on the schedule for quantum annealing as the entry point and thus concerns quantum annealers in contrast to most other techniques discussed so far that focus on algorithms for gate-based quantum computers.
Since we found only four techniques of this kind in the dataset, the information on this class is not sufficient to justify more precise subclasses.
We note, however, that there are techniques in this class that train ML models to generate annealing schedules (P07, P19) as well as techniques that rely on classical pre-processing (P52, P72-T2).
They can, thus, be seen as annealing equivalents to the techniques focusing on parameter initialization described above, likewise pre-computing or generating annealing schedules instead of initial variational parameters.\looseness=-1

\smallskip\noindent
\textbf{Warm-Starting via Circuit Synthesis.}
The last class identified in our dataset are techniques that utilize information classically derived from the targeted problem instance to influence the construction of variational circuits.
Therefore, they adapt the quantum algorithm at hand for a non-reflexive C2Q warm-start.
For example, the structure of quantum circuits can be derived from a classically approximated solution (P41) or the set of operators used for the circuit synthesis can be reduced based on classical approximations (P24).

\smallskip
The scheme shown in~\Cref{fig:results-classification-scheme} provides a general overview of different kinds of warm-starting techniques and shows how they relate to each other.
It aims to help quantum software engineers identify what kind of techniques may be applicable to their use cases, and, thus, facilitates the integration of suitable warm-starting techniques into their applications.
Moreover, researchers can utilize it as a basis for the exploration and comparison of existing and new warm-starting techniques in the quantum computing domain.
Furthermore, based on novel, emerging warm-starting techniques, the presented classification scheme can be extended with additional or refined classes and subclasses in the future.
\looseness=-1

%% file: figures/table_computational_problems.tex
\begin{table}[h]
    \centering
    \caption{Problems tackled with warm-started algorithms, with their number of occurrence and the identifiers of the related techniques as defined in \Cref{subsec:problems_algos_benefits}.}
    \begin{tabularx}{\linewidth}{p{5.25cm}c>{\small}X}
        \toprule
        \textbf{Problem} & \textbf{\#} & {\normalsize \textbf{Identifiers of Related Techniques}} \\ \hline
        Classification & 33 & P09-T1; P09-T2; P09-T3; P10; P11; P12; P18; P23; P28; P29; P30; P31; P32; P33; P35; P36; P37; P38; P40; P42; P45; P51; P53; P56; P57; P59; P60; P62; P65; P66; P70; P71; P78 \\
        Maximum Cut & 26 & P01; P03; P04; P05; P06; P13; P14; P15; P16; P17; P21; P25; P27; P34; P41; P44; P46; P47; P48; P49; P55-T1; P55-T2; P72-T1; P72-T2; P77; P79 \\
        Image Classification & 25 & P09-T1; P10; P11; P18; P28; P30; P31; P33; P35; P37; P38; P40; P42; P45; P51; P53; P56; P59; P60; P62; P65; P66; P70; P71; P78 \\
        Ground State Estimation & 8 & P03; P08-T1; P08-T2; P14; P24; P26; P27; P80 \\
        TFIM (Traverse Field Ising Model) & 4 & P58-T1; P58-T2; P76-T1; P76-T2 \\
        Ground State Preparation & 3 & P20; P58-T1; P58-T2 \\
        Molecular Hamiltonian & 2 & P50; P61 \\
        Quantum State Classification & 2 & P09-T2; P09-T3 \\
        QUBO (Quadratic Unconstrained Binary Optimization) & 2 & P52; P63 \\
        Sherrington-Kirkpatrick Ising Models & 2 & P17; P34 \\
        3-SAT & 1 & P19 \\
        Eigenvalue & 1 & P22 \\
        Encoding & 1 & P02 \\
        Graph Bisection & 1 & P80 \\
        Knapsack Problem & 1 & P74 \\
        MAX-2-SAT & 1 & P80 \\
        MAX-3-SAT & 1 & P75 \\
        Modularity Maximization & 1 & P69 \\
        Molecular Geometry Optimization & 1 & P39 \\
        Portfolio Optimization & 1 & P04 \\
        Prime Factorization & 1 & P07 \\
        Transmon Simulation  & 1 & P50 \\
        Variational State Preparation & 1 & P43 \\
        XXZ Spin Chain & 1 & P50 \\
        \bottomrule
    \end{tabularx}
    \label{tab:results-problems}
\end{table}

%% file: figures/table_quantum_algos.tex
\begin{table}[t]
    \centering
    \caption{Kinds of quantum algorithms involved in warm-starting techniques, with their number of occurrences in the dataset and the identifiers of the related techniques as defined in \Cref{subsec:problems_algos_benefits}.}
    \begin{tabularx}{\linewidth}{l>{\RaggedRight}p{4.85cm}c>{\small}X}
        \toprule
        \textbf{Abbr.} & \textbf{Algorithm} & \textbf{\#} & {\normalsize \textbf{Identifiers of Related Techniques}} \\ \hline
        QAOA & Quantum Approximate Optimization Algorithm & 35 & P01; P03; P04; P05; P06; P13; P14; P15; P16; P17; P20; P21; P25; P34; P41; P43; P44; P46; P47; P48; P49; P55-T1; P55-T2; P63; P64; P69; P72-T1; P73-T1; P73-T2; P74; P76-T1; P76-T2; P77; P79; P80 \\
        QNN & Quantum Neural Network & 30 & P02; P09-T1; P09-T2; P09-T3; P10; P11; P12; P18; P23; P28; P30; P31; P36; P37; P38; P39; P40; P42; P45; P51; P53; P56; P57; P59; P60; P62; P65; P70; P71; P78 \\
        VQE & Variational Quantum Eigensolver & 11 & P03; P17; P23; P24; P26; P27; P50; P54; P61; P68; P80 \\
        HVA & Hamiltonian Variational Ansatz & 4 & P08-T1; P08-T2; P58-T1; P58-T2 \\
        VQC & Variational Quantum Circuit & 4 & P33; P35; P38; P40 \\
        HEA & Hardware-Efficient Ansatz & 3 & P08-T1; P08-T2; P39 \\
        VQA & Variational Quantum Algorithm & 3 & P27; P29; P67 \\
        AQC & Adiabatic Quantum Algorithm & 2 & P07; P19 \\
        PQC & Parameterized Quantum Circuit & 3 & P03; P14; P32 \\
        QA & Quantum Annealing & 2 & P52; P72-T2 \\
        TQA & Trotterized Quantum Annealing & 1 & P13 \\
        QPE & Quantum Phase Estimation & 1 & P22 \\
        RBM & Restricted Boltzmann Machine & 1 & P66 \\
        QCBM & Quantum Circuit Born Machine & 1 & P68 \\
        CTQW & Continuous-Time Quantum Walk & 1 & P75 \\
        QBM & Quantum Boltzmann Machine & 1 & P39 \\
        \bottomrule
    \end{tabularx}
    \label{tab:results-algos}
\end{table}

%% file: figures/table_properties.tex
\begin{table}[]
    \centering
    \caption{Relevant properties of warm-starting techniques with a description of possible values and their number of occurrences (\#) in the dataset.}
    \begin{tabularx}{\linewidth}{@{}>{\RaggedRight}p{2.45cm}|p{4.7cm}>{\small}Xc@{}}
        \toprule
        \textbf{Property} & \textbf{Value} & {\normalsize \textbf{Description}} & \textbf{\#}  \\ \hline
        Direction & \textbullet\ C2Q & Classical-to-Quantum: A quantum algorithm is warm-started utilizing classically generated information. & 49 \\
        & \textbullet\ Q2Q & Quantum-to-Quantum: A quantum algorithm is warm-started utilizing information generated by a quantum algorithm. & 37 \\
        & \textbullet\ Q2C & Quantum-to-Classical: A classical algorithm is warm-started utilizing information generated by a quantum algorithm. & 2 \\ 
        \hline

        Quantum Device & \textbullet\ Annealer & A quantum annealer is used. & 5\\
        & \textbullet\ Gate-Based & A gate-based quantum computer is used. & 82 \\
        & \textbullet\ Both & Both kinds of devices are used. & 1 \\ 
        \hline

        Entry Point & \textbullet\ Encoding in initial quantum state & Information is encoded into the initial quantum state to conduct a warm-start. & 37\\
        & \textbullet\ Quantum circuit construction & The warm-start affects the circuit construction. & 2 \\ 
        & \textbullet\ Variational parameter initialization & Generating initial parameters for VQAs. & 36 \\ 
        & \textbullet\ Annealing schedule & Generating an annealing schedule. & 4 \\ 
        & \textbullet\ Optimization procedure & Enhancing the optimization procedure. & 7 \\ 
        & \textbullet\ Encoding in classical neural network & Encoding information into a classical neural network. & 1 \\ 
        & \textbullet\ Parameter initialization in classical neural network & Generating parameters to initialize a classical neural network. & 1 \\ 
        \hline

        Change of Quantum Algorithm & \textbullet\ Yes & A quantum algorithm is adapted for the warm-start, i.e., a quantum circuit in the gate-based scenario. & 14 \\
        & \textbullet\ No & No changes of quantum algorithms. & 74 \\ 
        \hline
                
        Reflexivity & \textbullet\ Reflexive & Information utilized to warm-start the targeted algorithm is generated by an instance of the same algorithm. & 34 \\
        & \textbullet\ Non-Reflexive & Information is generated differently. & 54 \\ 
        \hline

        Data Generation & \textbullet\ Only targeted problem instance required & Only the targeted problem instance is needed to conduct the warm-start, including in any pre-processing. & 35 \\
        & \textbullet\ Data from other problem instances required & Additional problem instances are solved to enable a warm-start. & 53 \\ 
        \hline

        Intermediary \newline Processing & \textbullet\ Machine Learning (ML) & Information is processed using ML to conduct a warm-start, e.g., to learn a model. & 40\\
        & \textbullet\ Quantum Machine Learning (QML) & Information processing by QML. & 5 \\
        & \textbullet\ None & No or only trivial (classical) processing. & 43 \\ 
        \bottomrule
    \end{tabularx}
    \label{tab:results-properties}
\end{table}

%% file: content/Limitations.tex

\section{Limitations and Threats to Validity}
\label{sec:limitations}
The type of study conducted here inherently suffers from certain limitations due to the research design.
In this section, we discuss possible limitations that pose threats to the validity of the results.\looseness=-1

\subsection{Selection Bias}
The first kind of limitations comes with the selection of sources for the literature search, the selection of search terms, the subsequent composition of search queries, and selection of relevant publications.
We considered six major publication databases for the database search.
However, relevant publications may simply not be listed in these publication databases.
Furthermore, although we included numerous alternative search terms in relatively open queries for the database search, we may have missed additional terms that also indicate warm-starting techniques in the quantum computing domain.
Therefore, the search queries composed with these search terms may have been too selective to find some relevant publications.
Both aforementioned limitations are mitigated by forward and backward snowballing to a certain extent, since snowballing is neither limited to a specific set of databases nor based on the composition of search queries.
However, forward snowballing in particular is again limited by the scope and particularities of the citation database employed for the task.
Nonetheless, we assume that Google Scholar's massive citation database, which was used for this task, includes the vast majority of relevant publications.
Further limitations come with the manual screening and filtering of publications, where different views of the authors come into play and render these tasks subjective.
While the general screening is less susceptible to subjectivity, as it only serves to determine whether a publication is related to the quantum computing domain, the selection of publications according to the inclusion and exclusion criteria is much more prone to different opinions.
We tried to mitigate this effect as much as possible by redundant processing and subsequent discussions among the authors about any contradictions.
Another minor limitation that comes with the selection criteria is the exclusion of works that do not provide sufficient information.
For example, very few publications, e.g. by~\citet{wang2021quantum}, had to be excluded since they did not provide sufficient details for the data extraction and classification, even though they were considered by the authors to fulfill the other selection criteria.
Also, a few closely related works on quantum algorithms for the purpose of transfer learning, predominantly several works by~\citet{he2020quantum}, were found not to fulfill the selection criteria and were excluded as their only relation to quantum computing is the fact that the techniques themselves, but neither source nor target algorithms, are implemented as quantum algorithms.
\looseness=-1

\subsection{Data Extraction}
The data extraction from the publications is another manual process potentially suffering from subjectivity.
Therefore, the extraction was cross-validated among the participating authors and reviewed by the first author for consistency.
Any contradictions were discussed until a consensus was reached, thereby mitigating the subjectivity in the data extraction.
A minor limitation of the study lies in the time-consuming processing of the publications and data extraction.
For example, more pre-prints may have been formally published after the bibliographical data from the pre-print had already been extracted.
However, we expect that this applies at most to a few works, which were nonetheless considered herein even if they may have been unjustly listed as pre-prints.
Moreover, the aggregation and analysis of the extracted data is a potential source for manual errors.
We addressed such potential errors by using spreadsheets and automating these steps with formulas.
Although data clustering could have been employed to group the warm-starting techniques based on their properties, we decided to identify the clusters manually, which avoids the problem of predetermining a specific number of clusters to be identified.
Additionally, the manual grouping allowed for immediate identification and treatment of outliers based on a consensus among the authors.
\looseness=-1

\subsection{Completeness}
Last, we note that the identified properties and classification scheme by nature make no claims of being complete, but rather document the current state-of-the-art research on the topic.
Researchers may find additional relevant properties of warm-starting techniques in the quantum computing domain in the future, particularly when new techniques have been proposed.
The presented classes of techniques could then be extended and further refined based on such properties.

%% file: content/RelatedWork.tex

\section{Related Work}
\label{sec:relatedwork}

SMSs are common research tools in software engineering.
Well-recognized guidelines for such studies have been introduced~\cite{petersen2008guidelines} and further improved~\cite{petersen2015guidelinesUpdate} by Petersen et al.
Also Kitchenham et al. provide relevant guidelines with a focus on literature search~\cite{kitchenham2007guidelinesSLR} as well as suggestions specifically for SMSs~\cite{kitchenham2012mapping}.
SMSs in software engineering focus on a wide range of topics, e.g., Function-as-a-Service platforms and tools~\cite{yussupov2019smsFaaS}, smart contract languages for blockchains~\cite{varela2021smsSmartContracts}, and natural language processing for requirements
engineering~\cite{zhao2021smsNLP}.
\looseness=-1

Despite an abundance of research on warm-starting in the fields of classical computing and algorithms (see \Cref{sec:background}), to the best of our knowledge, no secondary or tertiary research has been conducted on the topic.
\citet{Weigold2021_HybridPatterns} document warm-starting as a pattern for hybrid quantum algorithms.
They discuss warm-starting as a method of inducing a quantum-classic split in hybrid quantum algorithms, i.e., a composition of classical and quantum algorithms as seen in existing warm-starting techniques for QAOA and VQE~\cite{egger2021warm,tate2020bridging,barkoutsos2018quantum}.
Moreover, \citet{mari2020transfer} consider different scenarios of transfer learning for hybrid QNNs discussing warm-starting techniques primarily based on their direction property.
With our study, we continue this line of work by identifying, characterizing, and categorizing further research on warm-starting techniques in the quantum computing domain and deepening the general understanding of such techniques.
\looseness=-1

In addition to warm-starting, various other approaches have been proposed to address the limitations of NISQ devices, including, in particular, error mitigation techniques and improved queuing mechanisms.
On the one hand, error mitigation focuses on reducing the impact of errors in quantum computation and comprises two major approaches.
Gate error mitigation aims to compensate the expected gate errors by adapting the quantum circuit before execution~\cite{He2020zeronoise} and readout error mitigation aims to reduce the errors occurring specifically during the measurements of quantum states through classical post-processing~\cite{bravyi2021mitigating,beisel2022rem}.
On the other hand, recent queuing mechanisms aim to improve the execution of VQAs by providing support for repeated access to NISQ computers through hybrid quantum-classical runtime environments and prioritized access to the device~\cite{karalekas2020}, e.g., through \textit{Sessions}\footnote{\url{https://quantum-computing.ibm.com/lab/docs/iql/manage/systems/sessions/}} on IBM's quantum computing platform.
Although these general approaches address similar limitations of NISQ devices, we argue that they should be considered complementary to warm-starting techniques.
In contrast to error mitigation and queuing, warm-starting can (i) reduce the actual quantum resource requirements of algorithms and thus achieve significant speedups, e.g., by leveraging widely available classical computing resources, (ii) reduce data requirements, e.g., by utilizing pre-trained models, and (iii) embed performance guarantees of classical algorithms within quantum algorithms.

%% file: content/Conclusion.tex

\section{Conclusion}
\label{sec:conclusion}

With the study, we have created a map of state-of-the-art research on warm-starting techniques in the context of quantum algorithms.
Specifically, we have identified relevant properties of such techniques and built a classification scheme based on these properties.
With respect to our research question, our results show that 
(i) research on warm-starting techniques in the quantum computing domain is increasing rapidly and driven especially by academic researchers, often in collaborations with  other research institutions and the industry;
(ii) research on these techniques focuses on VQAs, especially QAOA and QNNs, and can be motivated mainly by speedups, reduced resource consumption, or increased solution quality;
(iii) the warm-starting techniques differ in a variety of specific properties and can be classified based on these properties in at least five main classes with ten distinct subclasses, where the encoding of information and parameter initialization play the most important role in realizing warm-starts.
We hope that the results presented in this work facilitate the usage and development of warm-starting techniques by researchers and quantum software engineers, as they provide an overview of available techniques, their properties, and the benefits of warm-starting.
\looseness=-1

In future work, we will focus on the identification of research gaps and aim to study warm-starting techniques in new scenarios that have not been considered yet in the literature.
In particular, we aim to further investigate the compatibility of different warm-starting techniques to apply multiple techniques simultaneously.
\mrev{Moreover, it would be interesting to further analyze some of the results of this study, e.g., to explore whether techniques employed by collaborators from the industry are applied to industry-specific applications and to evaluate the practical feasibility of warm-starting techniques by quantifying the improvements achieved with them.}
Furthermore, the identified properties and the classification scheme introduced in this work could serve as a basis for a decision support system for the application of warm-starting techniques in the context of quantum algorithms.
Therefore, we plan to explore this possibility by designing a decision support framework and appropriate tools.\looseness=-1

%% file: content/Appendix.tex

\section{Final Set of Selected Publications}
\label{sec:appendix-publications}
~\small{
\begin{longtblr}[
    caption={List of the final set of selected publications considered in this study.}, 
    label={tab:appendix-publications},
    note{(a)} = {Two publications are treated as one entity as described in~\Cref{subsec:combination}.}
    ]{
        colspec={lp{2.5cm}Xl},
        rowhead=1,
        rowsep=0pt,
        colsep=1pt,
    }
    \hline
    \textbf{ID} & \textbf{Authors} & \textbf{Publication Title} & \textbf{Year}  \\ 
    \hline
    P01\TblrNote{(a)} & Beaulieu and Pham & Evaluating performance of hybrid quantum optimization algorithms for MAXCUT Clustering using IBM runtime environment & 2021 \\
     & Beaulieu and Pham & Max-cut Clustering Utilizing Warm-Start QAOA and IBM Runtime & 2021 \\
    P02 & Chen and Yoo & Federated Quantum Machine Learning & 2021 \\
    P03 & Dborin et al. & Matrix Product State Pre-Training for Quantum Machine Learning & 2022 \\
    P04 & Egger et al. & Warm-starting quantum optimization & 2021 \\
    P05 & Galda et al. & Transferability of optimal QAOA parameters between random graphs & 2021 \\
    P06 & Jain et al. & Graph neural network initialisation of quantum approximate optimisation & 2021 \\
    P07 & Lin et al. & Hard instance learning for quantum adiabatic prime factorization & 2021 \\
    P08 & Liu et al. & A Parameter Initialization Method for Variational Quantum Algorithms to Mitigate Barren Plateaus Based on Transfer Learning & 2021 \\
    P09 & Mari et al. & Transfer learning in hybrid classical-quantum neural networks & 2020 \\
    P10 & Mittal and Dana & Gender Recognition from Facial Images using Hybrid Classical-Quantum Neural Network & 2020 \\
    P11 & Mogalapalli et al. & Classical{\textendash}Quantum Transfer Learning for Image Classification & 2021 \\
    P12 & Qi and Tejedor & Classical-to-Quantum Transfer Learning for Spoken Command Recognition Based on Quantum Neural Networks & 2022 \\
    P13 & Sack and Serbyn & Quantum annealing initialization of the quantum approximate optimization algorithm & 2021 \\
    P14 & Sauvage and Serbyn & FLIP: A flexible initializer for arbitrarily-sized parametrized quantum circuits & 2021 \\
    P15 & Tate et al. & Bridging Classical and Quantum with SDP initialized warm-starts for QAOA & 2020 \\
    P16 & Tate et al. & Classically-inspired Mixers for QAOA Beat Goemans-Williamson's Max-Cut at Low Circuit Depths & 2021 \\
    P17 & Verdon et al. & Learning to learn with quantum neural networks via classical neural networks & 2019 \\
    P18 & Alavi and Akhoundi & Hybrid Classical-Quantum method for Diabetic Foot Ulcer Classification & 2021 \\
    P19 & Chen et al. & Optimizing Quantum Annealing Schedules with Monte Carlo Tree Search enhanced with neural networks & 2022 \\
    P20 & Wauters et al. & Polynomial scaling of QAOA for ground-state preparation of the fully-connected p-spin ferromagnet & 2020 \\
    P21 & Shaydulin et al. & QAOAKit: A Toolkit for Reproducible Study, Application, and Verification of the QAOA & 2021 \\
    P22 & Guzman and Lacroix & Accessing ground state and excited states energies in many-body system after symmetry restoration using quantum computers & 2022 \\
    P23 & Tao et al. & LAWS: Look Around and Warm-Start Natural Gradient Descent for Quantum Neural Networks & 2022 \\
    P24 & Zhang et al. & Mutual information-assisted Adaptive Variational Quantum Eigensolver & 2021 \\
    P25 & Shaydulin et al. & Parameter Transfer for Quantum Approximate Optimization of Weighted MaxCut & 2022 \\
    P26 & Grimsley et al. & ADAPT-VQE is insensitive to rough parameter landscapes and barren plateaus & 2022 \\
    P27 & Ravi et al. & CAFQA: Clifford Ansatz For Quantum Accuracy & 2022 \\
    P28 & Umer et al. & An integrated framework for COVID-19 classification based on classical and quantum transfer learning from a chest radiograph & 2021 \\
    P29 & Yang et al. & When BERT Meets Quantum Temporal Convolution Learning for Text Classification in Heterogeneous Computing & 2022 \\
    P30 & Majumder et al. & Hybrid Classical-Quantum Deep Learning Models for Autonomous Vehicle Traffic Image Classification Under Adversarial Attack & 2021 \\
    P31 & Azevedo et al. & Quantum transfer learning for breast cancer detection & 2022 \\
    P32 & Jose and Simeone & Transfer Learning in Quantum Parametric Classifiers: An Information-Theoretic Generalization Analysis & 2022 \\
    P33 & Karur Mudugal Mathad et al. & Transfer Learning Using Variational Quantum Circuit & 2022 \\
    P34 & Chandarana et al. & Meta-Learning Digitized-Counterdiabatic Quantum Optimization & 2022 \\
    P35 & Kanimozhi et al. & Brain Tumor Recognition based on Classical to Quantum Transfer Learning & 2022 \\
    P36 & Koike-Akino et al. & Quantum Transfer Learning for Wi-Fi Sensing & 2022 \\
    P37 & Kumsetty et al. & TrashBox: Trash Detection and Classification using Quantum Transfer Learning & 2022 \\
    P38 & Ovalle-Magallanes et al. & Hybrid classical–quantum Convolutional Neural Network for stenosis detection in X-ray coronary angiography & 2022 \\
    P39 & Sbahi et al. & Provably efficient variational generative modeling of quantum many-body systems via quantum-probabilistic information geometry & 2022 \\
    P40 & Sridevi et al. & Quantum Transfer Learning for Diagnosis of Diabetic Retinopathy & 2022 \\
    P41 & Wurtz and Love & Classically Optimal Variational Quantum Algorithms & 2021 \\
    P42 & Acar and Yilmaz & COVID-19 detection on IBM quantum computer with classical-quantum transferlearning & 2020 \\
    P43 & Akshay et al. & Parameter Concentration in Quantum Approximate Optimization & 2021 \\
    P44 & Alam et al. & Accelerating Quantum Approximate Optimization Algorithm using Machine Learning & 2020 \\
    P45 & Amin et al. & A New Model for Brain Tumor Detection Using Ensemble Transfer Learning and Quantum Variational Classifier & 2022 \\
    P46 & Amosy et al. & Iterative-Free Quantum Approximate Optimization Algorithm Using Neural Networks & 2022 \\
    P47 & Boulebnane and Montanaro & Predicting parameters for the Quantum Approximate Optimization Algorithm for MAX-CUT from the infinite-size limit & 2021 \\
    P48 & Brandao et al. & For Fixed Control Parameters the Quantum Approximate Optimization Algorithm's Objective Function Value Concentrates for Typical Instances & 2018 \\
    P49 & Cain et al. & The QAOA gets stuck starting from a good classical string & 2022 \\
    P50 & Cervera-Lierta et al. & The Meta-Variational Quantum Eigensolver (Meta-VQE): Learning energy profiles of parameterized Hamiltonians for quantum simulation & 2020 \\
    P51 & Ciylan and Ciylan & Fake human face recognition with classical-quantum hybrid transfer learning & 2021 \\
    P52 & de Luis et al. & Steered quantum annealing: improving time efficiency with partial information & 2022 \\
    P53 & Gokhale et al. & Implementation of a quantum transfer learning approach to image splicing detection & 2020 \\
    P54 & Harwood et al. & Improving the Variational Quantum Eigensolver Using Variational Adiabatic Quantum Computing & 2022 \\
    P55\TblrNote{(a)} & Khairy et al. & Learning to Optimize Variational Quantum Circuits to Solve Combinatorial Problems & 2020 \\
     & Khairy et al. & Reinforcement-Learning-Based Variational Quantum Circuits Optimization for Combinatorial Problems & 2019 \\
    P56 & Kim et al. & Classical-to-quantum convolutional neural network transfer learning & 2022 \\
    P57 & Kulshrestha and Safro & BEINIT: Avoiding Barren Plateaus in Variational Quantum Algorithms & 2022 \\
    P58 & Mele et al. & Avoiding barren plateaus via transferability of smooth solutions in Hamiltonian Variational Ansatz & 2022 \\
    P59 & Mir et al. & Diabetic Retinopathy Detection Using Classical-Quantum Transfer Learning Approach and Probability Model & 2021 \\
    P60 & Mishra and Samanta & Quantum Transfer Learning Approach for Deepfake Detection & 2022 \\
    P61 & Mitarai et al. & Quadratic Clifford expansion for efficient benchmarking and initialization of variational quantum algorithms & 2020 \\
    P62 & Mogalapalli et al. & Trash classification using quantum transfer learning & 2022 \\
    P63 & Moussa et al. & Unsupervised strategies for identifying optimal parameters in Quantum Approximate Optimization Algorithm & 2022 \\
    P64 & Okada et al. & Systematic study on the dependence of the warm-start quantum approximate optimization algorithm on approximate solutions & 2022 \\
    P65 & Otgonbaatar et al. & Quantum Transfer Learning for Real-World, Small, and Large-Scale Datasets & 2022 \\
    P66 & Piat et al. & Image classification with quantum pre-training and auto-encoders & 2018 \\
    P67 & Rad et al. & Surviving The Barren Plateau in Variational Quantum Circuits with Bayesian Learning Initialization & 2022 \\
    P68 & Rudolph et al. & Synergy Between Quantum Circuits and Tensor Networks: Short-cutting the Race to Practical Quantum Advantage & 2022 \\
    P69 & Shaydulin et al. & Multistart Methods for Quantum Approximate Optimization & 2019 \\
    P70 & Skolik et al. & Layerwise learning for quantum neural networks & 2021 \\
    P71 & Soto-Paredes and Sulla-Torres & Hybrid Model of Quantum Transfer Learning to Classify Face Images with a {COVID}-19 Mask & 2021 \\
    P72 & Streif and Leib & Training the Quantum Approximate Optimization Algorithm without access to a Quantum Processing Unit & 2020 \\
    P73 & Truger et al. & Selection and Optimization of Hyperparameters in Warm-Started Quantum Optimization for the {MaxCut} Problem & 2022 \\
    P74 & van Dam et al. & Quantum Optimization Heuristics with an Application to Knapsack Problems & 2021 \\
    P75 & Wang & Classically-Boosted Quantum Optimization Algorithm & 2022 \\
    P76 & Wauters et al. & Reinforcement Learning assisted Quantum Optimization & 2020 \\
    P77 & Wurtz and Lykov & The fixed angle conjecture for QAOA on regular MaxCut graphs & 2021 \\
    P78 & Zhao et al. & Classical-Quantum Hybrid Transfer Learning for {COVID}-19 Classification & 2021 \\
    P79 & Lee et al. & Parameters Fixing Strategy for Quantum Approximate Optimization Algorithm & 2021 \\
    P80 & Wilson et al. & Optimizing quantum heuristics with meta-learning & 2021 \\
    \hline
\end{longtblr}
}